\documentclass[acmsmall,screen,nonacm,authorversion]{acmart}
\pdfoutput=1
 
\AtBeginDocument{%
  \providecommand\BibTeX{{%
    Bib\TeX}}}

\usepackage{amsmath,amssymb,amsfonts}

\usepackage{algorithm}
\usepackage{algorithmic}

\usepackage{graphicx}
\usepackage{textcomp}
\usepackage{xcolor}
\usepackage{amssymb}
\usepackage{dsfont}
\usepackage{multirow}
\usepackage{enumitem}
\usepackage{booktabs}
\usepackage{tcolorbox}
\usepackage{tabularx}
\usepackage{subfigure}
\usepackage{subcaption}
\usepackage{makecell}
\usepackage{tikz}
\usepackage{url}
\usepackage{balance}
\usepackage{utfsym}

\def\BibTeX{{\rm B\kern-.05em{\sc i\kern-.025em b}\kern-.08em
    T\kern-.1667em\lower.7ex\hbox{E}\kern-.125emX}}

\begin{document}

\title{MCGMark: An Encodable and Robust Online Watermark for Tracing LLM-Generated Malicious Code}
\fancyfoot{} 
        \renewcommand{\headrulewidth}{0pt} 
        \fancypagestyle{titlepage}{
        \fancyhf{} 
        \renewcommand{\headrulewidth}{0pt} 
        }

\author{Kaiwen Ning}
\affiliation{%
  \institution{Sun Yat-sen University and Peng Cheng Laboratory}
  \city{Guangdong}
  \country{China}}
\email{ningkw@mail2.sysu.edu.cn}

\author{Jiachi Chen}
\affiliation{%
  \institution{Sun Yat-sen University}
  \city{Zhuhai}
  \country{China}}
\email{chenjch86@mail.sysu.edu.cn}

\author{Qingyuan Zhong}
\affiliation{%
  \institution{Sun Yat-sen University}
  \city{Zhuhai}
  \country{China}}
\email{zhongqy39@mail2.sysu.edu.cn}

\author{Tao Zhang}
\affiliation{%
  \institution{Macau University of Science and Technology}
  \city{Macao}
  \country{China}}
\email{tazhang@must.edu.mo}

\author{Yanlin Wang}
\affiliation{%
  \institution{Sun Yat-sen University}
  \city{Zhuhai}
  \country{China}}
\email{wangylin36@mail.sysu.edu.cn}

\author{Wei Li}
\affiliation{%
  \institution{Sun Yat-sen University}
  \city{Zhuhai}
  \country{China}}
\email{liwei378@mail2.sysu.edu.cn}

\author{Jingwen Zhang}
\affiliation{%
  \institution{Sun Yat-sen University and Peng Cheng Laboratory}
  \city{Zhuhai}
  \country{China}}
\email{zhangjw273@mail2.sysu.edu.cn}

\author{Jianxing Yu}
\affiliation{%
  \institution{Sun Yat-sen University}
  \city{Zhuhai}
  \country{China}}
\email{yujx26@mail.sysu.edu.cn}

\author{Yuming Feng}
\affiliation{%
  \institution{Peng Cheng Laboratory}
  \city{Shenzhen}
  \country{China}}
\email{fengym@pcl.ac.cn}

\author{Weizhe Zhang}
\affiliation{%
  \institution{Harbin Institute of Technology and Peng Cheng Laboratory}
  \city{Guangdong}
  \country{China}}
\email{wzzhang@hit.edu.cn}

\author{Zibin Zheng}
\affiliation{%
  \institution{Sun Yat-sen University}
  \city{Zhuhai}
  \country{China}}
\email{zhzibin@mail.sysu.edu.cn}

\renewcommand{\shortauthors}{Ning et al.}

\begin{abstract}

With the advent of large language models (LLMs), numerous software service providers are developing LLMs tailored for code generation, such as CodeLlama. However, these models can be exploited by malicious developers to generate malicious code, posing severe threats to the software ecosystem. To address this issue, we first conducted an empirical study and built \textsc{MCGTest}, a dataset of $406$ prompts designed to elicit malicious code from LLMs. Leveraging this dataset, we propose \textsc{MCGMark}, a watermarking method to trace and attribute LLM-generated malicious code. \textsc{MCGMark} subtly embeds user-specific information into generated code by controlling the token selection process, ensuring the watermark is imperceptible. Additionally, \textsc{MCGMark} dynamically adjusts the token selection range to induce the LLM to favor high-probability tokens, thus ensuring code quality. Furthermore, by leveraging code structure, \textsc{MCGMark} avoids embedding watermarks into regions easily modified by attackers, such as comments and variable names, enhancing robustness against tampering. Experiments on several advanced LLMs show that \textsc{MCGMark} successfully embeds watermarks in approximately $85\%$ of cases, under the constraint of a $400$-token limit. Moreover, it maintains code quality and demonstrates strong resilience against common code modification. This approach offers a practical solution for tracing malicious code and mitigating the misuse of LLMs.

\end{abstract}


\maketitle

\section{Introduction}
\label{sec:introduction}

Code generation has become a crucial topic in software engineering~\cite{liu2023codegen4libs,tosem3}. It enables the automatic generation of code snippets from natural language requirements and significantly reduces manual coding efforts~\cite{Hao2024CodeGen,tosem2}. Recently, with the advent and development of large language models (LLMs), their potential in code-related tasks has been widely recognized~\cite{Du2023ClassEval,tosem1}. In response, Software Service Providers (SSP) are dedicating efforts to develop LLMs specifically tailored for code generation tasks, such as CodeLlama~\cite{roziere2023codellama} and DeepSeek-Coder~\cite{guo2024deepseek}.

However, despite their benefits, LLMs are also exploited for malicious purposes. Prior research~\cite{zhengju1,chen2024rmcbench,meta} reveals that malicious developers leverage LLMs to develop malware, such as spyware and ransomware. In addition, reports from organizations like CheckPoint~\cite{zhengju2} and CrowdStrike~\cite{zhengju3} highlight a growing trend of malicious software and cyberattacks facilitated by LLMs. Numerous cases and posts on technical forums further demonstrate the use of LLMs in generating harmful code~\cite{zhengju4,zhengju5,zhengju7,zhengju6}, posing significant risks to the software ecosystem.


Tracing malicious code generated by LLMs can effectively mitigate the abuse of LLMs. However, the existing methods, such as zero-shot detectors~\cite{mitchell2023detectgpt} and fine-tuning language model detectors~\cite{chen2023gpt}, have been proven to be ineffective in practical use~\cite {wu2023survey,yao2023survey}. For instance, OpenAI discontinued its classifier due to low accuracy (around $26\%$)~\cite{zhengju6}.

As an alternative, watermarking technology is considered a promising solution for tracing the origin of content generated by LLMs~\cite{fernandez2023three,takezawa2023necessary}. It embeds identifiable characteristics into the generated content, either explicitly or implicitly, to distinguish and attribute its origin~\cite{LiWWG23}. However, existing watermarking methods still face several key challenges when applied to tracing malicious code~\cite{li2023zhengju2,liu2023survey}. \textbf{(1) Traceability and Implicitness.} Current watermarking methods primarily focus on detecting whether a piece of content is generated by an LLM, but they overlook tracing the identity of the generator~\cite{lee2024zhengju3,liu2023zhengju4}. Moreover, most approaches add watermarks only after the content is generated by the LLM~\cite{chang2024postmark,yang2024srcmarker}, typically using lexical substitution, which makes them easier to identify and remove~\cite{fernandez2023three}. \textbf{(2) Ensuring code generation quality. }Mainstream implicit watermarking techniques are designed for natural language text, whereas code exhibits strong structural constraints and strict syntax requirements, making it challenging to embed watermarks without affecting functionality or quality~\cite{He2023llmcode}. \textbf{(3) Resistant to Tampering.} Since code elements like comments and variable names can be easily altered by attackers, watermarking methods should account for code structure to improve tamper resistance and robustness~\cite{TipirneniZR24}.

To address the aforementioned issues, in this paper, we propose \textsc{MCGMark}, a watermarking framework for tracing LLM-generated code. Our approach implicitly embeds encodable watermarks during the code generation process, considering the code's structure and ensuring the quality of the generated code. \textit{Firstly}, \textsc{MCGMark} implicitly embeds encodable watermarks by controlling the LLM’s token selection, ensuring the watermark is difficult to discern while reflecting the generator's identity. \textit{Secondly}, \textsc{MCGMark} dynamically obtains the probability distribution of candidate tokens and constrains the LLM’s selection to higher-probability tokens, thereby ensuring the quality of the watermarked code. \textit{Lastly}, we introduce a watermark skipping mechanism guided by code structure and syntax rules, allowing \textsc{MCGMark} to decide whether to embed watermarks in subsequent code elements during LLM code generation. This ensures that watermarks are not embedded in easily modifiable code elements, such as comments and variable names, thereby enhancing the robustness of the watermark. \textit{Additionally}, we conduct an empirical study on existing instances of malicious code. We collect $129$ real malicious code examples generated by LLMs and analyze $21,959$ malicious code repositories on GitHub. Based on the empirical study, we construct the first prompt dataset specifically designed for malicious code generation, comprising $406$ tasks to guide watermark design and assess its effectiveness.

\textsc{MCGMark} is designed as a plugin, decoupled from the LLM. During watermark embedding, it requires no additional models, data or tools. During watermark detection, \textsc{MCGMark} does not need to load LLMs. To implement \textsc{MCGMark}, the SSP only needs to adapt its token-matching rules to align with the specific LLM vocabulary. It is important to note that \textsc{MCGMark}, as a watermarking method, cannot eliminate the malicious nature of generated code or prevent its generation. Moreover, \textsc{MCGMark} applies watermarking to all generated code, regardless of its intent. Since benign code can potentially be repurposed to construct malicious software, MCGMark does not attempt to classify code as malicious or benign.

We apply \textsc{MCGMark} to three advanced LLMs to evaluate its effectiveness, while also introducing other baselines for a more comprehensive performance analysis. \textsc{MCGMark} embeds a $24$-bit watermark in $400$ tokens, achieving a watermark embedding success rate of about $85\%$ across different LLMs. Additionally, it outperforms other baselines in watermark detection success rate. Next, we assess the impact of \textsc{MCGMark} on code quality using the CodeBLEU~\cite{ren2020codebleu} and conduct a user study to further validate the results. The results demonstrate that \textsc{MCGMark} achieves significantly higher CodeBLEU scores than baseline methods, confirming its effectiveness in preserving code quality during watermark embedding. Furthermore, we evaluate \textsc{MCGMark} against $500$ program pairs and $1200$ modification attacks, demonstrating its effectiveness in resisting modification attacks. Finally, we analyze the impact of \textsc{MCGMark}’s hyperparameters and evaluate the time overhead of \textsc{MCGMark}.

In summary, this work contributes the following:
\begin{itemize}

    \item  We construct \textsc{MCGTest}, the first dataset for LLM-based malicious code generation, comprising $406$ prompts derived from real-world cases.
   
    \item We propose \textsc{MCGMark}, a robust and encodable watermarking scheme to trace LLM-generated code. \textsc{MCGMark} implicitly embeds user identity information in code generation, ensuring both code quality and robustness against watermark tampering.
    
    \item We evaluate \textsc{MCGMark} on multiple LLMs through comparative experiments. The results show that it successfully embeds a $24$-bit watermark with a success rate of approximately $85\%$ under a $400$ token output limit. Moreover, \textsc{MCGMark} demonstrates competitive performance in preserving code quality, resisting various attacks, and maintaining low time overhead, outperforming existing baseline methods in multiple aspects.

    \item We will release the source code of MCGMark and the related datasets after the paper is accepted to support further research.
\end{itemize}

\section{Background and Challenges}

\subsection{Code Generation of LLM}

Large Language Models (LLMs) are large-scale language models based on the Transformer architecture~\cite{bietti2024birth} and are trained on massive corpora, typically with billions of parameters or more~\cite{yang2024gpt4tools}. In recent research, LLMs have demonstrated impressive performance in code generation tasks~\cite{fan2023automated}
, which can significantly improve software development efficiency~\cite{liu2024your}. The performance of LLMs in code generation tasks has received extensive research attention~\cite{ShinWWSW24,liu2024your}. Moreover, some Software Service Providers (SSPs) have developed dedicated LLMs specifically designed for code generation, commonly referred to as Code LLMs. Currently, SSPs have developed numerous popular Code LLMs, such as Code Llama~\cite{roziere2023codellama}, DeepSeek-Coder~\cite{guo2024deepseek}, and StarCoder2~\cite{lozhkov2024starcoder}.

\begin{figure}[t]
	\centering{\includegraphics[scale=0.25]{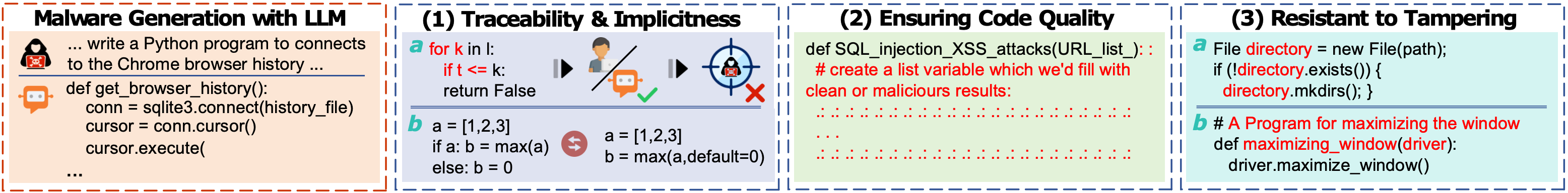}}
	\caption{The motivation example and challenges of design watermark against LLM-generated malicious code.}
	\label{Fig:mot}
\end{figure}

\subsection{Motivation}

LLMs are increasingly being exploited by malicious developers for generating malicious code~\cite{zhengju8,zhengju9,zhengju10}. Numerous instances have demonstrated the high efficacy of LLMs in producing harmful software~\cite{zhengju3,zhengju10,zhengju4,zhengju5,zhengju9}. Fig.~\ref{Fig:mot} illustrates a real-world example where an LLM was prompted to generate code for stealing browser history~\footnote{https://github.com/AI-Generated-Scripts/GPT-Malware}. This irresponsible utilization of LLMs for malicious code generation could pose a significant threat to the security of the software ecosystem.

Moreover, prior studies have demonstrated that LLMs can be easily misused to generate malicious code~\cite{chen2024rmcbench,lin2024malla}. For instance, RMCBench~\cite{chen2024rmcbench} evaluates the resistance of 11 representative LLMs against malicious code generation. The results show that the average refusal rate across all LLMs is only 28.71\%, and LLMs with varying generation capabilities can all be used to produce malicious code. Furthermore, malicious developers can employ instruction hijacking~\cite{qiang2023hijacking} and jailbreaking~\cite{niu2024jailbreaking} to further facilitate the generation of malicious code through LLMs. Therefore, it is imperative to design alternative approaches for LLMs to combat malicious code generation, with watermarking schemes emerging as one of the most promising solutions~\cite{liu2023survey}.

\subsection{Challenges}
\label{sec:motancha}

Designing watermarks to trace the generation of malicious code introduces several challenges.

\begin{itemize}[leftmargin=1em, itemsep=0pt, topsep=0pt, parsep=0pt, partopsep=0pt]
    \item \textbf{Traceability \& Implicitness.} The watermark should be accurately reflect the user's ID and implicit. Fig.\ref{Fig:mot}.(1).a shows an example of a watermark from the work~\cite{lee2024zhengju3}. Applying this watermark only indicates whether the code was generated by an LLM, failing to trace to a specific user. Fig.\ref{Fig:mot}. (1).b illustrates a pattern from the watermarking technique in the work~\cite{li2024resilient}, which adopts a post-processing watermarking strategy. This technique does not intervene during the code generation process. Instead, it modifies the code after generation, such as performing code transformation~\cite{yang2024srcmarker}. However, such approaches rely on predefined transformation patterns, which are inherently limited in applicability and cannot ensure compatibility across diverse code structures. In addition, watermarks based on fixed patterns tend to introduce noticeable artifacts, increasing the risk of being recognized and removed by malicious developers. Therefore, watermarking mechanisms should be designed to remain imperceptible while reliably encoding user-specific information.

    \item \textbf{Ensuring Code Generation Quality.} The embedding of watermarks must maintain the quality of the generated code. Fig.\ref{Fig:mot}.(2) shows an example of a watermark from the literature~\cite{Kirchenbauer2023watermark}. In this instance, watermark embedding significantly degraded the code quality, rendering the LLM-generated code unusable. In contrast to natural language, code is generally more structured and constrained by strict syntactic and semantic rules~\cite{He2023llmcode}. Some multi-encoding watermarking techniques attempt to mitigate quality degradation by leveraging the strong generative capabilities of LLMs~\cite{yoo2023robust,yoo2024advancing}. However, such approaches are not well-suited for code generation, where even slight modifications to the code may compromise functionality or correctness. Therefore, minimizing the impact of watermarking on code quality during generation remains a key challenge.

    \item \textbf{Resistant to Tampering.} Code elements, such as comments and variable names, can be altered without affecting the code's functionality. If a watermark is embedded in these elements, it can be easily altered or removed. Fig.\ref{Fig:mot}.(3).a illustrates a watermark example from the literature~\cite{li2023zhengju2}, where the watermark is added to variable names and can be easily modified. Fig.\ref{Fig:mot}. (3).b shows an example of a watermark from the literature~\cite{Kirchenbauer2023watermark}, where the watermark is added to comments and can also be easily disrupted. Therefore, the watermark needs to possess sufficient robustness to prevent it from being easily removed. However, online watermark embedding requires that code generation and watermark insertion occur simultaneously. Without access to the complete LLM output during embedding, the watermark must rely on incomplete context, making it difficult to ensure watermark robustness.
\end{itemize}

\section{\textsc{MCGTest:} A prompt dataset for LLM malicious code generation}
\label{sec:promptdataset}

In this section, we conduct an empirical study on real-world malicious code to help design our watermark. And we conduct \textsc{MCGTest}, a dataset of malicious code generation prompts that includes both actual instances of LLM-generated malicious code and potential scenarios.

\subsection{Data Collection}
\label{sec:3.1}

To thoroughly cover malicious code generation scenarios involving LLMs, we include both real-world examples and potential cases to ensure comprehensive coverage.

 \textbf{\textit{(Part 1)} Existing Instances Collection. }This involves gathering existing instances of malicious code generated by LLMs. We collect data from two major technical communities, GitHub~\cite{laiyuan1} and Stack Overflow~\cite{laiyuan2}, three literature databases, Google Scholar~\cite{laiyuan3}, arXiv~\cite{laiyuan5}, DBLP~\cite{laiyuan4}, and the Google search engine~\cite{laiyuan6}. We use the following four keywords for the above six platforms to collect results from January 2023 to March 2024: ``large language model malicious code," ``large language model malware," ``GPT malware," and ``GPT malicious code." In total, we collect $3,644$ results, including code repositories, papers, and articles.

 \textbf{\textit{(Part 2)} Potential Scenarios Collection.} This involves gathering possible scenarios for using LLMs to generate malicious code. For this part, we primarily collect repositories related to malicious code from GitHub. Using the keywords ``malicious code" and ``malware," we identify and collect $21,959$ malicious code repositories.

\subsection{Data Pre-processing}
\label{sec:3.2}

In this process, we describe the preprocessing of the collected data to extract malicious code instances, ensuring the relevance of the data for subsequent analysis.

\begin{figure}[t]
	\centering{\includegraphics[scale=0.23]{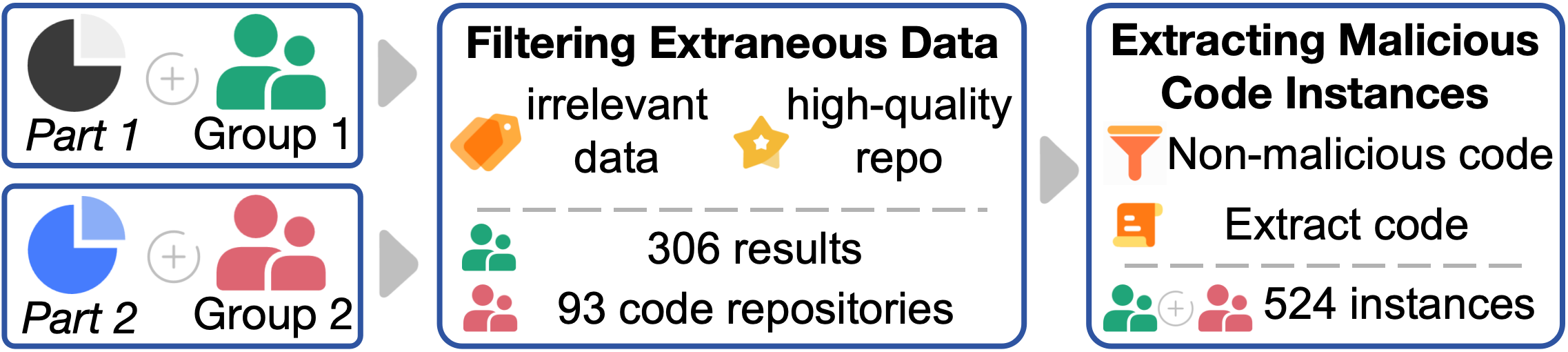}}
	\caption{The process of data pre-processing.}
	\label{Fig:preprocess}
\end{figure}

To identify both real instances and potential scenarios of LLM-generated malicious code, we design a two-step data preprocessing pipeline, as illustrated in Fig.~\ref{Fig:preprocess}. First, we remove irrelevant data, such as empty repositories. Then, we extract representative malicious code instances from the remaining data to analyze their functionality and construct the \textsc{MCGTest} dataset.

To ensure the reliability of this analysis. Four researchers with over four years of software development experience are assigned to filter and analyze the results. They are classified into two groups. Group 1 analyzes the data from \textit{Part 1}, while Group 2 is responsible for \textit{Part 2}. Members in each group work independently and then align conflicting results. Each group focuses only on their respective portion of the data, without interfering with each other.

 \textbf{Filtering Extraneous Data. }In this step, we primarily filter out results unrelated to malicious code. Group 1 removes irrelevant data of \textit{Part 1}, including empty code repositories, advertisements without substantial content, and web pages with only titles. Additionally, duplicate results and results unrelated to malicious code, such as evaluations or fixes using LLM for malicious code or security vulnerabilities of LLM itself, also be removed. Group 2 primarily selects high-quality malicious code repositories from \textit{Part 2}, excluding repositories unrelated to the topic or invalid repositories, such as malicious code detection tools or repositories that do not provide access to source code directly. Furthermore, to ensure an adequate number of collected malicious code instances, repositories with at least $200$ stars will be considered~\cite{sulun2024empirical}.  After alignment within the groups, Group 1 obtains a total of $306$ results, including $128$ literature references, five code repositories, and $173$ relevant web pages and articles. Group 2 collects $93$ code repositories.

 \textbf{Extracting Malicious Code Instances. }In this step, both groups are tasked with obtaining instances of malicious code generated by LLMs from the filtered results obtained in the previous stage. This forms the foundation of our LLM malicious code prompt dataset. Group 1 extracts descriptions of malicious code generated by LLMs from literature, news articles, and code repositories. They also identify and analyze malicious code snippets to determine their functionality. Group 2, on the other hand, selects usable malicious code functionality from the malicious code repositories. They exclude incomplete or ambiguous code and functions that are unrelated to the repository description or purpose, such as data visualization or graphical user interfaces. Additionally, to maintain the quality of collected instances, Group 2 members will exclude functions with fewer than five lines of code. After alignment within the groups, Group 1 and  Group 2 have selected $129$ and $395$ instances of malicious code, respectively.

In summary, we collected 524 malicious code instances, including LLM-generated samples and key open-source malicious code, covering potential generation scenarios.

\subsection{The Construction Process of Prompt Dataset}
\begin{figure}[t]
	\centering{\includegraphics[scale=0.9]{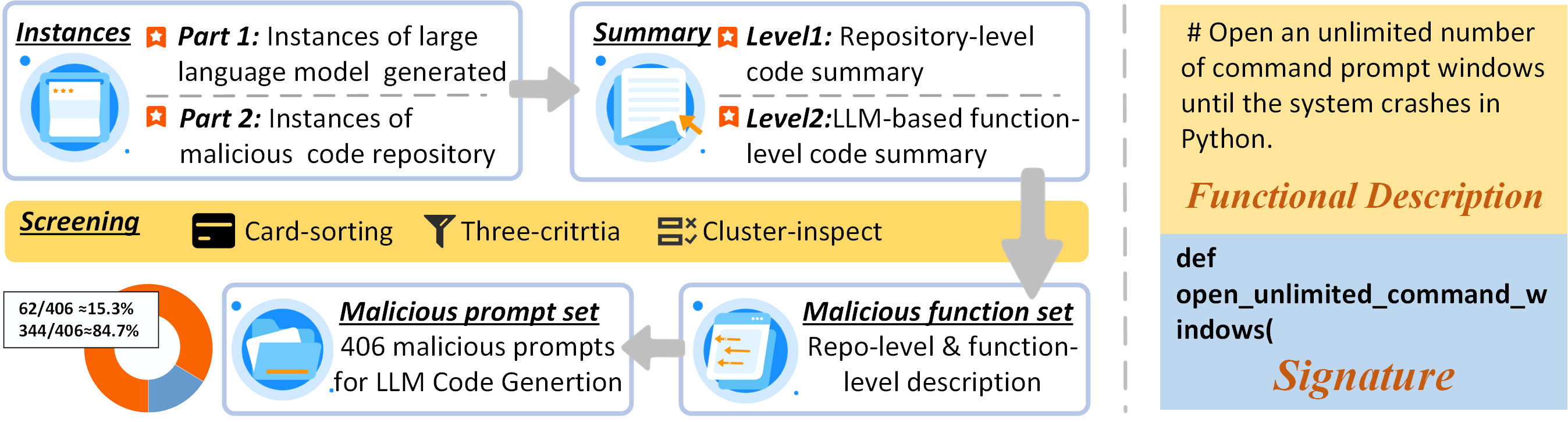}}
	\caption{The construction process and the prompt format of \textsc{MCGTest}.}
	\label{Fig:promptprocess}
\end{figure}

In this process, we construct the \textsc{MCGTest} prompt dataset for LLM-based malicious code generation using data collected in Section~\ref{sec:3.2}. As shown in Fig.~\ref{Fig:promptprocess}, the process involves three steps: (1) summarizing the functionality of malicious code instances; (2) filtering out redundant or unsuitable cases; and (3) creating prompts based on the remaining instances.

 \textbf{Summary of Malicious Code Functionality.} In this step, we aim to collect comprehensive information about malicious code's functionality, including both the overall intent of each instance and its specific malicious components. To achieve this, we establish a Malicious Function Set (MFS) to consolidates these functionalities. First, we summarize the functionality of each malicious code instance based on descriptions from their repositories or literature and add these summaries to the MFS. Next, we divide all instances into individual functions and use GPT-4~\cite{maniparambil2023enhancing}, an advanced LLM, to generate code summaries for these functions. Finally, we incorporate these summaries into the MFS. As a result, the MFS captures malicious behavior at both the instance and function levels, providing a comprehensive view of malicious code functionalities within our dataset.

 \textbf{Filtering Malicious Functionalities. }In this step, we filter malicious functionalities because not all of them are necessary in MFS. For example, copyright declarations or redundant functionalities across different functions. To ensure filtering accuracy, we employ a closed card sorting method to ensure the accuracy of results. Closed card sorting is one of the most efficient methods for organizing information into logical groups~\cite{ChenXLGLC22}. Two participants are involved, both have over four years of programming experience. The card title is code functionalities, and the description consists of code or descriptions from the code repository/literature. Both participants read and filter cards according to specified criteria, aligning their results. The overall filtering rules are as follows: 
\begin{itemize}[leftmargin=1em, itemsep=0pt, topsep=0pt, parsep=0pt, partopsep=0pt]
    \item Remove cards with duplicate semantic titles;
    \item Remove cards with ambiguous semantic titles, such as functions with unclear meanings;
    \item Remove cards with non-malicious semantic titles, such as copyright declaration functions.
\end{itemize}
Following these rules, we obtain a total of $406$ cards, comprising $72$ from \textit{Part 1} and $369$ from \textit{Part 2}. These $406$ cards correspond to $406$ distinct malicious code functionalities.

 \textbf{Creating the Malicious Prompt Dataset. }In this step, we construct the \textsc{MCGTest} dataset, which comprises 406 prompts derived from filtered cards describing malicious code functionality. Three participants, each with over four years of programming experience, collaborated on creating these prompts. Every prompt was initially drafted and reviewed by two participants, with a third participant resolving any disagreements. For the prompt format, we drew inspiration from MBPP and HumanEval~\cite{zheng2023codegeex}, two well-known datasets for evaluating LLM code generation performance. Each prompt consists of a function description and a function name, as illustrated in Fig.~\ref{Fig:promptprocess}. \textsc{MCGTest} is designed to be compatible with various LLMs. Typically, an LLM has multiple versions, differing mainly in parameter count and whether they are Base or Chat versions~\cite{He2023llmcode,chang2023survey}. Base versions are usually continuation models with limited human interaction capabilities, while Chat versions can engage in dialogue. \textsc{MCGTest}'s prompt format is compatible with both versions.

In summary, we construct $406$ malicious code generation tasks for \textsc{MCGTest}. These tasks include real instances of LLM-generated malicious code as well as prominent potential scenarios. 

\section{\textsc{MCGMark:}An Encodable and Robust watermark for LLM code generation}
\label{sec:watermark}

In this section, we introduce \textsc{MCGMark}, a method for embedding encodable watermarks during LLM code generation. 

\subsection{Overview}

\begin{figure}[t]
	\centering{\includegraphics[scale=0.315]{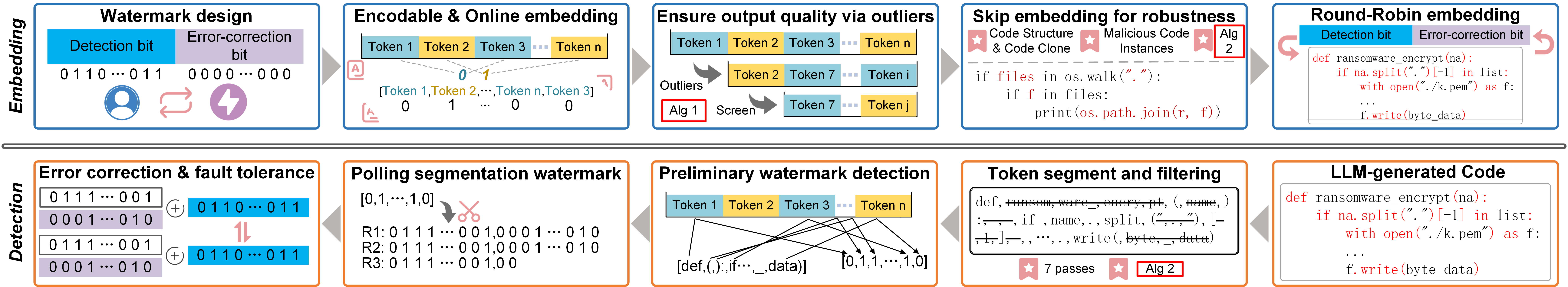}}
	\caption{The overview of watermark embedding and detection.}
	\label{Fig:watermarkprocess}
\end{figure}

Fig.~\ref{Fig:watermarkprocess} outlines the watermark embedding and detection process of \textsc{MCGMark}. Each process comprises five steps. 

 \textbf{Watermark embedding process. } \textsc{MCGMark} first initializes the watermark based on the user's ID. The watermark consists of detection bits and error-correction bits, which collectively represent the user's ID~\textit{(Section~\ref{sec:watermark-part2}.1 and Section~\ref{sec:watermark-part4}.1)}. Subsequently, \textsc{MCGMark} partitions the LLM's vocabulary and embeds multi-encoded watermarks by controlling the LLM's token selection~\textit{(Section~\ref{sec:watermark-part1})}. To mitigate the impact on code generation quality during watermark embedding, \textsc{MCGMark} then processes probability outliers in the vocabulary to ensure the LLM selects high-probability tokens, and updates the error-correction bit information~\textit{(Section~\ref{sec:watermark-part2})}. Next, to enhance watermark robustness, we design a watermark skipping strategy for \textsc{MCGMark} based on code structure and syntax. \textsc{MCGMark} implements this strategy based on the generated code elements, ensuring that watermarks are not added to easily modifiable code elements. Finally, as code generation progresses, watermarks are embedded in a round-robin fashion to further enhance their robustness~\textit{(Section~\ref{sec:watermark-part3})}.

It is important to note that \textsc{MCGMark} operates independently of the LLM’s internal generation process. It does not interrupt or roll back token generation, nor does it rely on any additional models or external databases during embedding. Moreover, \textsc{MCGMark} is fully decoupled from the LLM architecture and does not participate in its neural computations. As a result, techniques such as fine-tuning, distillation, or prompt engineering have no effect on the watermarking process.

 \textbf{Watermark detection process.} \textsc{MCGMark} first tokenizes the code into a sequence of tokens with the tokenizer. Next, \textsc{MCGMark} removes tokens that would have been skipped during the embedding process, based on the employed skip strategy. Subsequently, \textsc{MCGMark} partitions the vocabulary and examines the vocabulary membership of tokens in the sequence to recover the watermark information. Since the watermark is embedded in multiple rounds, \textsc{MCGMark} then trims the recovered watermark. Finally, \textsc{MCGMark} reconstructs the user's ID from the multiple segments of the trimmed watermark~\textit{(Section~\ref{sec:watermark-part4}.2)}.

It is important to note that \textsc{MCGMark} does not require simulating the LLM’s code generation process or accessing the LLM during detection. Only the tokenizer is required, which is used to split the code into tokens. This allows \textsc{MCGMark} to identify the vocabulary group each token belongs to and recover the embedded watermark information.

\subsection{Encodable Watermark Embedding}
\label{sec:watermark-part1}

In this process, \textsc{MCGMark} embeds encodable watermarks during the code generation by controlling token selection.

 \textbf{LLM Code Generation Process. }In a typical LLM code generation process, the LLM maintains a token-level vocabulary $V=\left\{v_0, v_1, \cdots, v_n\right\}$, typically comprising approximately $3.2 \times 10^4$ tokens (e.g., the DeepSeek-Coder-6.7b utilizes a vocabulary of $32,022$ tokens)~\cite{kandpal2023large}. When a prompt $R$ is input into the model $M$, it first employs a tokenizer to segment $R$ into token-level components, $R \Rightarrow T=\left\{t_0, t_1, \cdots , t_m\right\}$. Subsequently, $M$ computes the probability distribution $P_1=\left\{P_0^{\prime}, P_1^{\prime}, \cdots, P_n^{\prime}\right\}$ for all tokens in the vocabulary based on $T$. Then, $M$ selects the highest probability token $v_i^{\prime}(0 \leqslant i \leqslant n)$ from the vocabulary as the generated token and appends it to the prompt $R$. This process transforms the prompt $R$ into $R^{\prime}=\left\{t_0, t_1, \cdots , t_m, v_i^{\prime}\right\}$. $R^{\prime}$ is then fed back into model $M$, and this process iterates until a predetermined generation length $L$ is reached. The final output of the model is represented as $R^{(L)}=\left\{t_0, t_1, \cdots , t_m, v_i^{\prime}, v_i^{\prime \prime}, \cdots, v_i^{(L)}\right\}$, where $\left\{v_i^{\prime}, v_i^{\prime \prime} , \cdots , v_i^{(L)}\right\}$ constitutes the generated code~\cite{LiSCLZ23}.

 \textbf{Watermark Embedding. }Inspired by the study~\cite{Kirchenbauer2023watermark}, \textsc{MCGMark} encodes binary information by dividing the LLM’s vocabulary into two parts and sampling tokens from these parts based on the user’s ID. The vocabulary division is performed using a pseudo-random partitioning process on hash seeds, which is dynamically adjusted according to specific rules. This approach preserves the random characteristics of the vocabulary during embedding while allowing the randomness to be accurately reproduced during detection~\cite{hu2023unbiased,christ2024undetectable}.

\textsc{MCGMark} incorporates watermarking by modifying the vocabulary $V=\left\{v_0, v_1, \cdots, v_n\right\}$. \textsc{MCGMark} first generates a random set $D$ ($0<|D| \leqslant|V|$) with a hash value $H$. The elements in $D$ are unique, increasing integers representing vocabulary positions. They define the selected vocabulary $\mathbb{A}$, with the rest forming $\mathbb{B}$. \textsc{MCGMark} adjusts the LLM’s probability distribution to constrain token selection to $\mathbb{A}$. To reduce reliance on $H$ in partitioning, it applies a pseudo-random augmentation to $H$. This approach ensures that $H$ remains variable while maintaining it reproducible. As all generated tokens stem from randomly chosen vocabulary subsets, the LLM-generated code inherently differs from manually written code. The probability of complete overlap between manually-generated and model-generated code is merely $\frac{1}{2^L}$~\cite{Kirchenbauer2023watermark}. This low probability ensures the effectiveness of the code watermark.

 \textbf{Encoding Watermark.} \textsc{MCGMark} encodes the watermark to represent user-specific data. Specifically, \textsc{MCGMark} achieves encoding watermark embedding by modifying the probabilities of the LLM vocabulary. For embedding watermark $w_w$ into token $v_i^{(v)}$, \textsc{MCGMark} controls the LLM's selection as follows:
\begin{enumerate}[leftmargin=2em, itemsep=0pt, topsep=0pt, parsep=0pt, partopsep=0pt]
\item If the watermark bit is $1$, select a token from $\mathbb{A}$.
\item If the watermark bit is $0$, select a token from $\mathbb{B}$.
\end{enumerate}
To ensure watermark correctness, \textsc{MCGMark} guarantee that the model selects elements from the predefined vocabulary. This requires that the token with the highest probability in the modified vocabulary is present in the selected vocabulary. After generating element $v_i^{(v-1)}$, \textsc{MCGMark} obtains the probabilities of all tokens in the current vocabulary and calculates: $P_{\text {gap }}=\max \left\{P_{v}\right\}-\min \left\{P_{v}\right\}~. $ \textsc{MCGMark} then adds $P_{\text{gap}}$ to all elements in the selected vocabulary to ensure this vocabulary contains the highest probability token. 

Thus, \textsc{MCGMark} completes encodable watermark embedding in code generation.

\subsection{Ensuring Watermarked Code Quality}
\label{sec:watermark-part2}

In this process, \textsc{MCGMark} maintains the quality of the generated code by guiding the LLM to select from high-quality token candidates.

\begin{algorithm}[t]
\footnotesize
\caption{Preserving LLM Code Generation Quality through Outlier Management } 
\label{Alg:Alg1}

\begin{algorithmic}[1]
\STATE \textbf{Input}: Prompt: $R$, watermark: $W$, the number of new tokens: $L$, threshold for $P_{\text {dis }}$: $Thr\_P_{\text {dis }}$, a hash key $H$.
\FOR{$l = 0,1,\cdots,L$}
\STATE Obtain $V=\left\{v_0, v_1, \cdots, v_n\right\}$ and $P_{l}=\left\{P_0^{l}, P_1^{l}, \cdots, P_n^{l}\right\}$.
\STATE Calculate the outliers of $P_{l}$.
\IF{$F_{\text {upper }} \neq \emptyset $}
\STATE \textbf{Case 1:} $|F_{\text {upper }}| = 1$,$D^{'} = D \cup F_{\text {upper }}$.
\STATE \textbf{Case 2:} $|F_{\text {upper }}| \geq 2$, $D^{'} = D \cup F_{\text {upper }}[0: \lceil \frac{F_{\text {upper }}}{2}\rceil]$ with $H$. 
\ENDIF

\STATE Sample $v_i^{(l)}$ from $V$.
\IF{$v_i^{(l)} \in F_{\text{upper}} ~\&~ v_i^{(l)} \notin D$}
\STATE Setting the watermark's error-correction bit to $1$.
\ENDIF
\ENDFOR
\end{algorithmic}
\end{algorithm}

 \textbf{Code Quality Enhancement with Outliers.} To minimize the impact of watermarks on LLM code generation quality, \textsc{MCGMark} develops a watermark quality enhancement algorithm based on the probability outlier of the vocabulary, as shown in~\texttt{Algorithm~\ref{Alg:Alg1}}. This algorithm addresses potential issues arising from vocabulary partitioning, which may lead to incorrect selection of deterministic tokens and propagate errors in subsequent code generation. Leveraging the powerful generation capability of LLM, \textsc{MCGMark} implements the following strategy to ensure code generation quality during watermark embedding.

\begin{enumerate}[leftmargin=2em, itemsep=0pt, topsep=0pt, parsep=0pt, partopsep=0pt]
\item Before generating token $v_i^{(v)}$, \textsc{MCGMark} analyzes the probabilities $P$ to identify upper outliers—tokens with significantly higher probabilities.
\item If no upper outliers exist, \textsc{MCGMark} proceeds with standard watermark embedding.
\item If upper outliers are present and only for a single outlier, \textsc{MCGMark} includes it into the selected vocabulary.
\item For multiple outliers, \textsc{MCGMark} randomly selects half with $H$ and includes them in the selected vocabulary.
\end{enumerate}

However, outliers can affect the accuracy of watermark detection. During detection, we can only analyze the code and cannot obtain the real-time probability distribution of the vocabulary. This limitation aligns with real-world scenarios. To address this, we include error-correction bits in the watermark to recover the watermark information. When outliers impact the watermark information, \textsc{MCGMark} sets the error-correction bit to $1$; otherwise, it is set to $0$. Once the watermark detection bits are fully embedded, the error-correction bits are also generated. Subsequently, the error-correction bits are embedded into the code. During the embedding of error-correction bits, the watermark is not influenced by outliers.

\textbf{Outliers Detection.} \textsc{MCGMark} uses the Inter-Quartile Range (IQR) method, based on boxplot statistics~\cite{bondarenko2024quantizable}, to detect outliers in token probability distributions. IQR measures the spread of the middle $50\%$ of data by computing the difference between the third ($Q_3$) and first ($Q_1$) quartiles~\cite{vinutha2018detection}. This method is well-suited for \textsc{MCGMark} due to its robustness to extreme values and its effectiveness in non-normally distributed data~\cite{powell2023fort, yang2019outlier}. We define the upper whisker as Equ.~(\ref{equ:OUTLIER}), where $S$ is a scaling factor.
 
 \vspace{-1.5em}
 \begin{equation}
 \label{equ:OUTLIER}
 F_{\text{upper}} = Q_3 + S \cdot IQR = (S+1) \cdot Q_3 - S \cdot Q_1.
 \end{equation}
\vspace{-1.5em}

By detecting outliers in the LLM’s vocabulary during code generation, we are able to identify and handle tokens that are critical for maintaining code quality. This mechanism helps avoid selecting low-quality tokens that could compromise the readability or functionality of the code during watermark embedding.

In summary, after this step, \textsc{MCGMark} achieves multi-encoding watermark embedding during LLM code generation while maintaining code quality.

\subsection{Enhancing Watermark Robustness}
\label{sec:watermark-part3}

In this process, we provide a detailed description of \textsc{MCGMark}'s robustness scheme, designed based on the code structure. This scheme enables watermarks to withstand typical code modifications attempted by malicious developers. We delineate the adversarial scenarios and present a comprehensive overview of the design process.

 \textbf{Adversarial Scenarios.} Once malicious developers become aware of the possible presence of a watermark, they may attempt to tamper with the code. To avoid breaking its functionality, such developers, especially those with limited experience, tend to modify easily changeable elements, such as comments or variable names. Consequently, if the watermark is placed on easily modifiable elements, such as variable names, the watermark becomes highly susceptible to being rendered ineffective.

 \textbf{The Overall of Watermark Robustness Enhancement. }This step introduces the robustness enhancement strategy of \textsc{MCGMark} to address adversarial scenarios. To improve the resilience of watermarking in LLM-generated code, \textsc{MCGMark} avoids embedding watermarks in code regions that are easily modified or removed. However, since watermark embedding must be synchronized with code generation, \textsc{MCGMark} needs to make real-time decisions on whether to embed a watermark for each token. To enable this, the embedding process is divided into two stages.
\begin{enumerate}[leftmargin=2em, itemsep=0pt, topsep=0pt, parsep=0pt, partopsep=0pt]
    \item \textsc{MCGMark} identifies code elements to be excluded from watermark embedding by defining skipping rules grounded in code structure, code cloning patterns, and insights from real-world malicious code.
    \item \textsc{MCGMark} decides whether to embed a watermark for the next token, $v_i^{(l+1)} (l \in [0, L - 1])$, based on the tokens already generated by the model, $R^{(l)}=\left\{t_0, t_1, \cdots, t_m, v_i^{\prime}, v_i^{\prime \prime}, \cdots, v_i^{(l)} \right\}$. Since the tokens in the LLM vocabulary do not strictly adhere to human grammar rules, especially in the case of Code LLM, such as ``:(" or ``])", we design a watermark skipping scheme for \textsc{MCGMark} based on the grammar rules of the code and the LLM vocabulary.
\end{enumerate}

It is worth noting that due to significant differences in code structure among different programming languages, and since Python is currently one of the most commonly used languages by malicious developers~\cite{pythongongji1}, \textsc{MCGMark} focuses solely on Python language.

 \textbf{Watermark Skipping Rules. }In this step, we describe the element selection criteria of \textsc{MCGMark} for watermark skipping. To design watermark skipping rules, it is crucial to identify which elements in Python code are easily modifiable without affecting code usability. Our analysis focuses on three key perspectives:
\begin{itemize}[leftmargin=1em, itemsep=0pt, topsep=0pt, parsep=0pt, partopsep=0pt]
    \item \textit{(Code Structure.)} Python primarily consists of the following elements~\cite{schwarz2020overview}: indentation, keywords, identifiers, statements, expressions, functions, objects, object methods, types, numbers, operators, comments, exception handling, input/output, and blank lines. Previous research indicates that modifications to certain elements have minimal impact on code execution quality: identifiers (including variable names, function names, and class names), comments, output statements, numerical values, and blank lines~\cite{Zhang2024ZL23, funabiki2022study}.
    \item \textit{(Code Clone.)} Code clone detection focuses on identifying plagiarized code using similarity metrics~\cite{xu2024dsfm,li2024prism}, typically categorized into four levels: exact, lexical, syntactical, and semantic. Exact clones differ only in whitespace, layout, or comments; lexical clones use different identifiers but retain structure; syntactical clones modify statements while preserving structure; semantic clones perform the same function with different syntax. Recent studies show that detecting exact and lexical clones is feasible~\cite{codeclone1,codeclone2,codeclone6,codeclone7,codeclone8,codeclone9,codeclone10}, indicating that modifications like whitespace, layout, comments, and identifiers are relatively easy~\cite{codeclone3,codeclone4,codeclone5}.
    \item \textit{(Malicious Code Instances.)} We analyze the existing instances of malicious LLM code in Section~\ref{sec:3.2}. We observe that assignments, comparisons, and parenthetical elements, besides comments, output, and identifiers, are also readily modifiable in these instances. Thus, when designing the watermark, we must avoid embedding it in elements related to these operations. 
\end{itemize}

\begin{table}[t]
\small
\centering
\caption{Code Elements Excluded from Watermarking}\label{tab:EazyModi}
\scalebox{0.9}{
\begin{tabular}{l||c}
\hline \textbf{Perspective} & \textbf{Elements in code that are susceptible to modification} \\
\hline Code Structure & identifiers, comments, output, numbers, blank lines \\
\hline Code Clone & exact clone features, lexical clone features \\
\hline Malicious Code Instances & comments, output, identifiers, assignments, comparisons \\
\hline
\end{tabular}
}
\end{table}

In summary, the watermark is not embedded in code elements listed in Table~\ref{tab:EazyModi}.

 \textbf{Watermark Skipping Pattern. }In this step, we define watermark skipping patterns in \textsc{MCGMark}, guided by skipping rules. Since LLM-generated tokens are irreversible and the watermark embedding process is synchronized with the token generation, \textsc{MCGMark} cannot wait for the model to generate elements in Table~\ref{tab:EazyModi} before skipping the watermark. The embedding modifies the distribution of $P_{l}=\left\{P_0^{l}, P_1^{l}, \cdots, P_n^{l}\right\}(l \in [0, L])$, influencing the LLM's decision-making. Therefore, watermark embedding must be controlled before generating elements in Table~\ref{tab:EazyModi}. Thus, \textsc{MCGMark} decides on watermark embedding for the next token based on the already generated code, considering that vocabulary tokens are irregular and may not correspond directly to code elements.

\begin{algorithm}[t]
\footnotesize
\caption{Enhancing the Robustness of Watermark via Code Structure and Syntax}
\label{Alg:Alg2}

\begin{algorithmic}[1]
\STATE \textbf{Input}: Existing tokens: $R^{(l)}=\left\{t_0, t_1, \cdots , t_m, v_i^{\prime}, v_i^{\prime \prime}, \cdots, v_i^{(l)}\right\}$ , watermark to be embedded $w_X$, set $\widehat{A,B,C,D}$.   
\IF{$l \leq L$}
\STATE Check $v_i^{(l)}$ only whitespace and rollback or keep the $X$. $\# \textbf{Pattern 5}$
\IF{not $\mathcal{LOCK}$}   
\IF{$v_i^{(l)} \in \left\{ \widehat{A} \cup \widehat{B} \cup \widehat{C} \cup \widehat{D} \right\}$}
\STATE $ \mathcal{LOCK} \Leftarrow 1$.  \quad \quad  \quad  \quad   \quad  \quad \quad  \quad $\# \textbf{Pattern 6}$
\STATE Rollback and skipping watermark information based on different patterns triggered by $v_i^{(l)}$ and update $w_X$. $\# \textbf{Pattern 1, 2, 3, 4}$
\STATE $V = V$, \textbf{break}.
\ELSE  
\IF{$X \leq x$} 
\STATE Take set $D$ or $V \cap D$ corresponding to $w_X$, to $V$, $X = X+1$.
\ELSE
\STATE X = 0, Iterative embedding the watermark. \quad $\# \textbf{Pattern 7}$
\ENDIF
\ENDIF
\ENDIF
\STATE Based on the effectiveness of Pattern to determine $ \mathcal{LOCK} \Leftarrow 0$.
\STATE $V = V$, \textbf{break}.
\ENDIF
\end{algorithmic}
\end{algorithm}

Seven patterns are designed for skipping watermark embedding during LLM code generation.
\begin{itemize}[leftmargin=1em, itemsep=0pt, topsep=0pt, parsep=0pt, partopsep=0pt]
    \item \textit{(Pattern 1.)} If \( v_i^{(l)} (l \in [0, L]) \) in \( \widehat{A} = \left\{ \text{def, class, print, pprint, int, float, str, for, while, if, elif} \right\} \), subsequent tokens are not watermarked until a token containing \( '\backslash n' \) appears. This is because elements in \( \widehat{A} \) are often followed by identifiers or output, so \textsc{MCGMark} do not watermark subsequent content until \( v_i^{(l)} \) is \( '\backslash n' \).
    \item \textit{(Pattern 2.)} If \( v_i^{(l)} \) belongs to the set \( \widehat{B} = \left\{ (, [, ', ", \{ \right\} \), then no watermark is applied to subsequent tokens until a matching symbol is encountered. This is because elements in set \( \widehat{B} \) are often followed by tokens containing identifiers, values, and other easily modifiable elements. Hence, \textsc{MCGMark} does not apply a watermark to the content inside parentheses or quotation marks. No processing is performed if a pair of matching symbols, such as ``$($",``$)$", appears within a single token.
    \item \textit{(Pattern 3.)} If \( v_i^{(l)} \) is in the set \( \widehat{C} = \left\{ =, ==, \#, >, <, \geq, \leq, \neq \right\} \), representing numerical comparisons, assignments, and comment symbols, no watermark is applied to subsequent tokens until a token containing \( '\backslash n' \) is encountered. Additionally, we need to roll back the watermark position. Except for \( '\#' \), which requires rolling back by 1 position, the rollback distance for other watermark elements is determined by the difference between the current token's watermark position and the closest \( '\backslash n' \)-containing token's watermark position. This approach ensures that: (a.) Numerical comparison and assignment symbols, often surrounded by identifiers, avoid watermarking to preserve the integrity of the entire line. Rolling back ensures modifications or deletions of identifiers adjacent to these symbols do not affect the watermark. (b.) Comments, including \( '\#' \), are easily modified or deleted and thus should not be watermarked.
    \item \textit{(Pattern 4.)} If \( v_i^{(l)} \) is in the set \( \widehat{D} = \left\{ ``````, '', ``` \right\} \), representing multi-line comments, no watermark is applied to subsequent tokens until the same element reappears. Additionally, the watermark is rolled back by $1$ position.
    \item \textit{(Pattern 5.)} If \( v_i^{(l)} \) consists solely of whitespace characters like \( '\backslash t' \) or \( '\backslash n' \), it is necessary to check if it contains watermark information. If so, the watermark should be rolled back by $1$ position. Otherwise, no action is taken.
    \item \textit{(Pattern 6.)} If one Pattern is active, conflicting Patterns are blocked from triggering. However, if conditions in \( v_i^{(l)} \) satisfy the triggering criteria of two Patterns simultaneously, only the first Pattern in sequence will be triggered.
    \item \textit{(Pattern 7.)} Once all watermark bits are embedded, but tokens are still being generated, \textsc{MCGMark} continues embedding to reinforce the watermark.
\end{itemize}

It is important to note that the aforementioned rollbacks refer only to the rollback of watermark information, not the rollback of tokens generated by the LLM. When watermark information is embedded into code elements that are prone to modification or removal, the watermark needs to roll back to ensure it remains intact and avoids being altered.

 \textbf{Watermark Skipping Process. }In this step, we describe \textsc{MCGMark}'s watermark skip decision process based on the established skip rules and patterns. \texttt{Algorithm}~\ref{Alg:Alg2} delineates the execution process of watermark skipping. After obtaining the sequence $R^{(l)}=\left\{t_0, t_1, \cdots , t_m, v_i^{\prime}, v_i^{\prime \prime}, \cdots, v_i^{(l)}\right\}$, \textsc{MCGMark} first verify if $l \leq L$, where $L$ represents the maximum output token limit of the LLM. If this condition is met, \textsc{MCGMark} proceeds with the watermarking process; otherwise, \textsc{MCGMark} terminates. \textsc{MCGMark} then evaluates $v_i^{(l)}$ against several conditions: if it consists solely of whitespace characters, \textsc{MCGMark} triggers \textit{Pattern 5}; \textsc{MCGMark} checks for any active patterns based on \textit{Pattern 6} which would preclude watermarking the next token; and \textsc{MCGMark} determines if $v_i^{(l)} \in \left\{ \widehat{A} \cup \widehat{B} \cup \widehat{C} \cup \widehat{D} \right\}$, which would trigger the corresponding \textit{Pattern (1, 2, 3, or 4)} along with \textit{Pattern 6 }to prevent multi-pattern conflicts. If no pattern is triggered, \textsc{MCGMark} selects a word from vocabulary $\mathbb{A}$ or $\mathbb{B}$ based on the watermark information. Upon complete embedding of all watermark information, \textsc{MCGMark} triggers \textit{Pattern 7} to iterate and incorporate additional watermark data. \texttt{Algorithm}~\ref{Alg:Alg2} ensures judicious watermark embedding while preserving code integrity and functionality, accounting for various code elements and potential pattern conflicts.

In summary, \textsc{MCGMark} embeds watermarks by encoding the code generator’s identity while preserving code quality and enhancing robustness. To illustrate how the components interact, the complete embedding process is presented in \texttt{Algorithm~\ref{Alg:Alg3}}. When the current token contains `\texttt{\textbackslash n}', watermark embedding starts from the next token. For each token requiring watermark embedding, the vocabulary is randomly split based on a hash value, and \texttt{Algorithm~\ref{Alg:Alg2}} determines whether to skip embedding or roll back the watermark. If embedding proceeds, \texttt{Algorithm~\ref{Alg:Alg1}} identifies outliers in the current vocabulary. A sampled outlier is then checked for inclusion in the intended partition: if absent, the tolerance bit is set to 1; otherwise, to 0. The outlier is sampled and used to generate the next token. When rollback is needed, the hash value is updated using the last valid token. If embedding is skipped, the vocabulary remains unpartitioned.

 \begin{algorithm}[t]
\footnotesize
\caption{The overview of \textsc{MCGMark} embedding.}
\label{Alg:Alg3}
 
\begin{algorithmic}[1]
\STATE \textbf{Input}: Prompt: $R$, watermark: $W$, number of tokens: $L$, threshold $Thr\_P_{\text{dis}}$, hash key $H$, set $\widehat{A,B,C,D}$.  
\FOR{$l = 0,1,\cdots,L$}
    \IF{current token contains `\texttt{\textbackslash n}'}
        \STATE Start watermark embedding from the next token.
        \STATE Randomly partition the vocabulary into two parts using $H$.
        \STATE Call \texttt{Algorithm 2} to check for skipping or rolling back watermark embedding.
        \IF{embedding is required}
            \STATE Call \texttt{Algorithm 1} to obtain outliers of the current vocabulary.
            \STATE \textbf{If} sampled outlier $\notin$ partition: set tolerance bit $= 1$, \textbf{else}: set tolerance bit $= 0$.
            \STATE Sample outlier and generate the token.
        \ELSIF{rollback is required}
            \STATE Update $H$ to hash of the last valid token.
        \ELSE
            \STATE Do not partition the vocabulary.
        \ENDIF
    \ENDIF
\ENDFOR
\end{algorithmic}
\end{algorithm}

\subsection{Design and Detection of Watermark}
\label{sec:watermark-part4}

In this process, we design watermark patterns for \textsc{MCGMark} to enhance watermark detection success rates and describe a lightweight watermark detection procedure.

 \textbf{Watermark Design.} This step delineates the watermark format design for \textsc{MCGMark}. \textsc{MCGMark} employs a dual-component watermark design comprising Detection Bits and Error Correction Bits. Detection Bits primarily encode user information for traceability, while Error Correction Bits facilitate the recovery of potentially erroneous information in the Detection Bits. Although \texttt{Algorithm~\ref{Alg:Alg1}} ensures LLM code generation quality, setting a low outlier threshold (small $S$ value) may result in excessive outliers, potentially impacting the LLM's word selection. For instance, the LLM might be compelled to select words from vocabulary $\mathbb{B}$ instead of $\mathbb{A}$ as dictated by the watermark information bit, due to outlier presence. This scenario could lead to errors in specific watermark bits. To mitigate this issue, \textsc{MCGMark} introduces Error Correction Bits to restore information in the Detection Bits and generates watermarks in the Detection Bits without outlier influence. Furthermore, Detection Bits and Error Correction Bits are designed with equal length. This design effectively addresses the trade-off between maintaining code quality and preserving watermark integrity, enhancing the overall robustness of \textsc{MCGMark}'s watermarking strategy.

\begin{algorithm}[t]
\footnotesize
\caption{The overview of \textsc{MCGMark} detecting.}
\label{Alg:Alg4}

\begin{algorithmic}[1]
\STATE \textbf{Input}: Code, a hash key $H$, set $\widehat{A,B,C,D}$.   
\STATE Tokenize the code into a sequence of tokens.
\STATE Call \texttt{Algorithm 2} to remove specific tokens.
\FOR{each token in the sequence}
    \IF{token contains `\texttt{\textbackslash n}'}
        \STATE Start watermark detection from the next token.
        \FOR{each subsequent token}
            \STATE Partition the vocabulary using $H$, verify token existence, and extract binary information.
            \STATE Update $H$ based on the last valid token.
        \ENDFOR
        \STATE Split the watermark information by bit length.
        \IF{bit length $<$ watermark length}
            \STATE Report detection failure.
        \ELSE
            \STATE \textbf{If} multiple rounds exist and conflict, report failure; \textbf{else} output watermark.
        \ENDIF
    \ENDIF
\ENDFOR
\end{algorithmic}
\end{algorithm}

It is important to note that the Error Correction Bits in \textsc{MCGMark} differ from the traditional concept of error correction codes in network protocols. In \textsc{MCGMark}, the Error Correction Bits are an essential component of the watermark itself, rather than an auxiliary feature. They serve as a core functionality of the watermarking mechanism.

 \textbf{Watermark Detection. }This step describes the lightweight watermark detection process for \textsc{MCGMark}. In this process, \textsc{MCGMark} requires the code to be detect, the LLM's vocabulary, tokenizer, \texttt{Algorithm~\ref{Alg:Alg1}}, and \texttt{Algorithm~\ref{Alg:Alg2}}. There is no need to load any LLM. Given the malicious code, \textsc{MCGMark} first tokenizes the code using the tokenizer. Next, it applies the seven patterns and \texttt{Algorithm~\ref{Alg:Alg2}} to remove elements where watermark embedding can be skipped, resulting in a sequence of code elements. Then, the hash value $H$ in \texttt{Algorithm~\ref{Alg:Alg1}} is used to partition the vocabulary. The code element sequence is then traversed, and elements are categorized into the corresponding vocabulary parts, producing a sequence of $0$s and $1$s. \textsc{MCGMark} subsequently segments this sequence according to the watermark length. Each segment provides detection bits and error-correction bits, which are used to obtain the user's identity information using the following formula:

\vspace{-1.5em}
 \begin{equation}
 \label{equ:detect}
 \text{Detection Bits} \oplus (\text{1}~\&~\text{Error Correction Bits})~.
 \end{equation}
\vspace{-1.5em}

Due to round-robin embedding, multiple watermark segments may be extracted. Consistent results from at least two rounds help identify the malicious code generator, improving fault tolerance.

The detailed watermark detection process of \textsc{MCGMark} is outlined in \texttt{Algorithm~\ref{Alg:Alg4}}. Use a tokenizer to divide the code into tokens and call \texttt{Algorithm~\ref{Alg:Alg2}} to remove specific tokens. The algorithm then iterates through all tokens; when a token contains `\texttt{\textbackslash n}’, detection begins from the next token. For each token to be checked, a hash value is used to partition the vocabulary, determine the token’s group, and extract a corresponding bit. The hash is updated based on the last valid token. Extracted bits are then grouped by watermark length: if insufficient bits are obtained, detection fails. Otherwise, the bits are organized into rounds. If multiple rounds conflict, detection also fails; if not, the watermark is returned. If only one round is present, it is returned directly.

\section{Experiments}

We evaluate \textsc{MCGMark} in this section based on the watermark design requirements analyzed in Section~\ref{sec:motancha}. Specifically, we focus on the following research questions: 
\begin{itemize}
    \item \textbf{RQ1.} What are the watermark embedding and detection success rates of \textsc{MCGMark}?
    \item \textbf{RQ2.} How does \textsc{MCGMark} affect the quality of generated code?
    \item \textbf{RQ3.} How robust is \textsc{MCGMark} in withstanding adversarial scenarios?
    \item \textbf{RQ4.} What factors influence the successful watermark embedding in \textsc{MCGMark}?
    \item \textbf{RQ5.} What is the time overhead of \textsc{MCGMark}?
\end{itemize}

For RQ1, we evaluate the watermark embedding success rate and detection success rate of \textsc{MCGMark} across different LLMs, while comparing it with other state-of-the-art watermarking methods. For RQ2, we compare the code generation quality of \textsc{MCGMark} and its baselines with the CodeBLEU~\cite{ren2020codebleu}. Additionally, we conduct a user study to assess the impact of watermark embedding on the quality of generated code.  For RQ3, we evaluate the robustness of \textsc{MCGMark} and its baselines under adversarial scenarios. Furthermore, for RQ4, we investigate the influence of key parameters on the embedding success rate of \textsc{MCGMark} and examined the efficiency of watermark generation. Finally, in RQ5, we investigate the impact of \textsc{MCGMark} on time overhead by using different LLMs and comparing against the baselines.

\subsection{Experiment Setup}

In this part, we present the experimental setup and the baseline methods used for comparison.

 \textbf{LLMs:} We evaluate our watermarking strategy on three state-of-the-art LLMs: DeepSeek-Coder-6.7b-instruct~\cite{guo2024deepseek}, StarCoder 2-7b-hf~\cite{lozhkov2024starcoder}, and CodeLlama-7b-hf~\cite{codellama2023}. These LLMs were selected because they are open-source, which allows us to access their vocabularies and implement \textsc{MCGMark}. Furthermore, they have demonstrated strong performance across multiple benchmarks~\cite{guo2024deepseek,ShinWWSW24}.

 \textbf{Parameters of \textsc{MCGMark}:} We set the number of LLM maximum output token, $L$, to $400$. Additionally, we set $\lceil \frac{|D|}{|V|}\rceil = 0.5$. The total length of the watermark, $X$, is set to $24$ bits. And the watermark information is randomly generated. $Q3$ is set to $0.75$, and $Q1$ is set to $0.25$~\cite{bondarenko2024quantizable}.

 \textbf{Evaluation Dataset:} We evaluate \textsc{MCGMark} using \textsc{MCGTest} and CodeBLEU~\cite{ren2020codebleu}. \textsc{MCGTest} consists of $406$ malicious code prompts, including real instances generated by LLMs and malicious code collected from high-quality open-source repositories. CodeBLEU is an evaluation framework designed for code generation tasks. It integrates traditional n-gram matching, syntax matching, and semantic matching to measure code generation quality comprehensively.

 \textbf{Baselines:} 
We introduce three state-of-the-art LLM watermarking techniques as baselines for comparison: WLLM~\cite{Kirchenbauer2023watermark}, MPAC~\cite{yoo2024advancing}, and PostMark~\cite{chang2024postmark}. WLLM embeds zero-bit watermarks by partitioning the LLM vocabulary into red and green subsets, ensuring the model selects tokens exclusively from the green subset. This approach focuses solely on identifying whether code is generated by an LLM. MPAC extends WLLM on embedding multi-bit watermarks by controlling token selection between vocabularies. Finally, PostMark is a post-processing watermarking method that uses a private vocabulary substitution table to indicate whether the content is LLM-generated. Like WLLM, PostMark is also a zero-bit watermarking scheme.

 \textbf{Parameters of baselines:} For the parameter configurations of the three baselines, we follow the default settings as specified in their original works. Specifically, for WLLM, $\gamma$ was set to $0.25$ and $\delta$ to $2$. To ensure fairness, WLLM’s generated token count was limited to $400$, and it utilized the same hash function as \textsc{MCGMark}. For MPAC, while maintaining its default settings, the watermark bit length was set to $24$, and the token count was also limited to $400$.

 \textbf{Implementation:} We implement \textsc{MCGMark} in Python. All experiments are conducted on a workstation with $128$ CPU cores and $8$ $\times$ NVIDIA A800 (80G) GPUs.

\subsection{RQ1: Effectiveness of \textsc{MCGMark}}

To answer RQ1, we first evaluate the watermark embedding success rate and detection success rate of \textsc{MCGMark} across different LLMs. This evaluation not only assesses the effectiveness of watermark embedding in LLMs but also demonstrates the adaptability of \textsc{MCGMark} to various models. Additionally, we compare the performance of different watermarking schemes under identical settings to further highlight the capabilities of \textsc{MCGMark}.

 \textbf{Watermark Embedding Success Rate. } In this part, we apply \textsc{MCGMark} to different LLMs and tested the success rate of watermark embedding using the \textsc{MCGTest} dataset. Watermark embedding is attempted three times. As shown in Table~\ref{table:llmcompare}, the average watermark embedding success rate across the three LLMs was 85.2\%. Specifically, \textsc{MCGMark} achieved the highest embedding success rate with DeepSeek-Coder at 88.9\%, while the success rates for CodeLlama and StarCoder-2 were 79.6\% and 87.2\%, respectively. These results demonstrate that \textsc{MCGMark} is adaptable to different LLMs, aligning with the theoretical findings reported by Kirchenbauer et al~\cite{kirchenbauer2023reliability}. Furthermore, the results validate the effectiveness of \textsc{MCGMark} in watermark embedding.

\begin{table}[h]
\centering
\small
\caption{Watermark Embedding Success Rate of \textsc{MCGMark} with various LLM.}\label{table:llmcompare}
\scalebox{0.9}{
\begin{tabular}{l||c|c}
\hline \textbf{LLM} &\textbf{Model} & \textbf{Sucess Rate.(\%)}  \\
\hline 

Deepseek-Coder & 6.7b-instruct & 88.9 \\
StarCoder-2 & 7b-hf & 87.2 \\
CodeLlama & 7b-hf & 79.6 \\
\hline
Average & / & 85.2 \\

\hline

\end{tabular}
}
\end{table}

Subsequently, to further explore the effectiveness of \textsc{MCGMark}, we analyze the $45$ instances where \textsc{MCGMark} failed to embed watermarks on DeepSeek-Coder. We find that $17$ tasks failed due to the generated code being too short to embed the watermark. Additionally, $25$ tasks failed because the generated code triggered numerous traits, resulting in a high number of assignment statements and comments, which hindered watermark embedding. Another three prompts reject the malicious code generation request, resulting in empty content. Higher folding watermark encoding could potentially solve the issue of shortcode generation failures, though this is not the focus of this paper. The failures related to code structure primarily stemmed from our restriction of LLMs to generate a maximum of $400$ tokens. In practical applications, LLMs typically generate $2048$ to $4096$ tokens or even more (e.g., DeepSeek-Coder-6.7b can generate up to 64K~\cite{guo2024deepseek}). We retest the $25$ tasks that fail due to code structure with a maximum length setting of $2048$ tokens and successfully embed watermarks in $21$ of them, achieving an embedding success rate of $84\%$. Therefore, under conditions where token numbers are unrestricted, \textsc{MCGMark}'s watermark embedding success rate could reach approximately $94.1\%$ across $406$ tasks.

 \textbf{Watermark Detecting Success Rate.} In this part, we evaluate the detection success rate of \textsc{MCGMark} and compared it with other watermarking strategies. To ensure fairness, we analyze the performance of the DeepSeek-Coder-6.7b-instruct model with different watermarking strategies on the \textsc{MCGTest} dataset. The results are presented in Table~\ref{table:rq1watermarkcompare}. As shown, \textsc{MCGMark} achieved the highest watermark detection success rate. Both WLLM and MPAC also demonstrated relatively high detection success rates. However, the detection success rate of PostMark, a post-processing watermarking method, was significantly lower. This can be attributed to PostMark’s reliance on the substitution model’s capability and the adaptability of its maintained substitution table.

\begin{table}[h]
\centering
\small
\caption{Watermark detecting Success Rate of various watermark.}\label{table:rq1watermarkcompare}
\scalebox{0.9}{
\begin{tabular}{l||c|c|c|c}
\hline \textbf{Watermark} &\textbf{In-process}  & \centering\textbf{Multibit} &\textbf{Code-aware}& \textbf{Detect Rate.(\%)}  \\
\hline 

WLLM & \usym{1F5F8} & \usym{2613}& \usym{2613} & 84.2 \\
PostMark & \usym{2613} & \usym{2613}& \usym{2613} & 21.4 \\
MPAC & \usym{1F5F8} & \usym{1F5F8}& \usym{2613} & 87.5\\
MCGMark & \usym{1F5F8} & \usym{1F5F8}& \usym{1F5F8} & 86.9 \\
\hline

\end{tabular}
}
\end{table}

 \textbf{Analysis of Detection Failures.} In this part, we first analyze the scenarios where \textsc{MCGMark} failed in watermark detection. Subsequently, we examine the impact of tolerance bits on the watermark detection success rate. For the first study, Out of the $361$ tasks where \textsc{MCGMark} successfully embed watermarks, $353$ watermarks are detected successfully, resulting in a detection success rate of approximately $97.8\%$. In contrast, WLLM does not support false positive rate checking. The eight instances where \textsc{MCGMark} failed to detect watermarks can be attributed to discrepancies in token splitting by the tokenizer. This issue leads to errors when verifying tokens against the vocabulary, returning incorrect results. This limitation is inherent to SSP, and we cannot improve the detection success rate by modifying \textsc{MCGMark}.

For the second part, we conduct a detailed evaluation of the impact of tolerance bits on watermark detection. Specifically, we assess the detection success rate of watermarks without applying tolerance bits in 361 cases where watermark embedding was successful. We observe that without error-correcting bits, the watermark is rarely detected successfully. This outcome can be attributed to two main reasons. First, the inherent randomness in vocabulary partitioning makes it difficult to ensure that outlier tokens consistently fall into the vocabulary group aligned with the watermark encoding. Second, the total watermark length is 24 bits, and under this setting, the theoretical probability of successful detection without error correction is only $0.19\%$. These results further highlight the necessity of incorporating error-correcting bits in our watermark design.

\begin{tcolorbox}[title = {Answer to RQ 1:}] 
\small

\textsc{MCGMark} achieved a watermark embedding success rate of over 85\% and a watermark detection success rate of 97.8\%. Compared to other watermarking methods, it demonstrated superior performance. Moreover, \textsc{MCGMark} is model-agnostic and does not rely on specific LLMs.
\end{tcolorbox}

\subsection{RQ2: Impact on LLM Code Generation}

To address RQ2, we first evaluate the quality of code generated under different watermarks with CodeBLEU~\cite{ren2020codebleu}, a widely adopted framework for assessing code quality. Additionally, we conduct a user study to further investigate the impact of \textsc{MCGMark} on the quality of code generation.

 \textbf{CodeBLEU Result.} We assess how different watermarking strategies affect code quality using CodeBLEU, comparing results to code generated without watermarking. Inspired by previous work~\cite{zhang2024watermarking,guan2024codeip}, we evaluate the impact of different watermarking methods on code quality by comparing CodeBLEU scores before and after watermark embedding. In this evaluation, a higher CodeBLEU score indicates a smaller impact of the watermark on the LLM. To ensure fairness, we address a specific behavior of PostMark. Since PostMark sometimes converts code entirely into plain text during watermark embedding—an inherent characteristic of its design—we assign a score of 0 to such cases. For all successfully watermarked code segments, we compute CodeBLEU scores before and after embedding and report the average score. The results are presented in Table~\ref{table:rq2watermarkcompare}.

In Table~\ref{table:rq2watermarkcompare}, the second column indicates whether the watermark is multi-bit. The third column shows whether the watermark embedding or detection process depends on external databases or requires loading additional models. As shown in Table~\ref{table:rq2watermarkcompare}, all watermarking strategies, including \textsc{MCGMark}, result in a decline in the code generation quality of LLMs. \textsc{MCGMark} achieved higher scores, indicating a smaller impact of the watermark on code generation. Furthermore, compared to other watermarking strategies, \textsc{MCGMark} supports embedding a greater amount of multi-bit information while having a less pronounced impact on code quality.

\begin{table}[h]
\centering
\small
\caption{CodeBLEU results of various watermark.}\label{table:rq2watermarkcompare}
\scalebox{0.9}{
\begin{tabular}{l||c|c|c}
\hline \textbf{Watermark} &\textbf{Multibit(digits)}  &\textbf{No external dependencies} & \textbf{Core}  \\
\hline 
WLLM  & \usym{2613} &\usym{1F5F8} & 0.19 \\
PostMark & \usym{2613} & \usym{2613} & 0.16 \\
MPAC & \usym{1F5F8} & \usym{1F5F8} & 0.21\\
MCGMark & \usym{1F5F8} & \usym{1F5F8} & 0.27 \\
\hline

\end{tabular}
}
\end{table}

Additionally, some watermarking strategies, such as PostMark, rely on external databases or models, requiring additional resources to be loaded during watermark embedding and detection. In contrast, \textsc{MCGMark} operates independently of such external dependencies.

 \textbf{User Study.} We further conduct a user study to evaluate the impact of \textsc{MCGMark} on the quality of LLM-generated code. We primarily assess the code generation quality of the LLM to explore the impact of \textsc{MCGMark} on the code generation quality. In this part, we randomly select $50$ tasks from the $346$ successfully embedded tasks of \textsc{MCGTest}. Furthermore, we obtain $100$ code segments, with $50$ generated by the model using our watermarking strategy and $50$ without. We invite $10$ developers with at least $4$ years of development experience (excluding co-authors), including $6$ Ph.D. students, $2$ undergraduate students, and $2$ software engineers specializing in computer-related fields, to participate in our evaluation. We randomly shuffle the order of the $50$ code pairs and further shuffle the code order within each pair. We ask the developers to identify the code they believe contains a watermark for each code pair. We collect a total of $1000$ valid responses and organize the accuracy of these $1000$ responses, as shown in Table~\ref{table:rq2dis}. 


\begin{table}[h]

\centering
\small
\caption{The results of the user study on distinguishing watermark and unwatermarked code.}\label{table:rq2dis}

\begin{minipage}[t]{\textwidth}
\centering
\scalebox{0.9}{
\begin{tabular}{l|c|c|c|c|c}
\hline
\multicolumn{2}{l||}{participants} & 1 & 2 & 3 & 4 \\
\hline
\multicolumn{2}{l||}{Correct Number / Success Rate} & 22/44\% & 27/54\%& 21/42\% & 26/52\%\\
\hline
\end{tabular}
}
\end{minipage}
\hfill
\vspace{2pt} \\
\begin{minipage}[t]{\textwidth}
\centering
\scalebox{0.9}{
\begin{tabular}{c|c|c|c|c|c||c}
\hline
5 & 6 & 7 & 8 & 9 & 10 & \textbf{Avg} \\
\hline
30/60\% & 19/38\%& 21/42\% & 24/48\% & 21/42\% & 28/56\% & \textbf{23.9/47.8\%}\\
\hline
\end{tabular}
}
\end{minipage}

\end{table}

The recognition accuracy of $10$ participants for $500$ pairs of watermark/unwatermark code, totaling $1000$ segments, is $47.8\%$, which is close to random sampling. Moreover, the independent recognition rates of the $10$ participants, excluding the highest value (participant 5) and the lowest value (participant 6), fluctuate between $40\%$ to $50\%$. Furthermore, all these rates fall within the $95\%$ confidence interval ($[21.972, 27.828]$). So, we can conclude that experienced practitioners with long-term development experience cannot correctly distinguish between watermark and unwatermark codes. This also demonstrates the stealthiness of our watermark and further confirms that the impact of \textsc{MCGMark} on code quality can be considered negligible.

\begin{tcolorbox}[title = {Answer to RQ 2:}] 
\small

\textsc{MCGMark} preserves higher code generation quality compared to baseline methods while embedding watermarks, without noticeably impacting the normal functionality of the LLM.
\end{tcolorbox}

\subsection{RQ3: Resistance to Tampering}
\label{sec:Exper-robust}

To address RQ3, we first evaluate the robustness of watermarks against eight types of attacks. Subsequently, we conduct a detailed analysis of the results for \textsc{MCGMark}.

 \textbf{Robustness Test.} In this part, our main focus is on the robustness of the watermark in Section~\ref{sec:watermark-part3}. Based on the literature~\cite{reopen2022}, we primarily consider the following two types of attacks, as shown in Table~\ref{table:attack}. These $8$ attack types cover the majority of modifications that typically employ against code watermarks~\cite{codeclone4,codeclone7}. These eight types of attacks do not compromise the functionality of the code, making them more representative of the behavior of malicious developers with low coding proficiency. We evaluate the performance of $50$ successfully watermarked in \textsc{MCGTest} codes under $8$ attacks, particularly whether these attacks affect the embedded watermark elements. For each attack, we conduct three attack instances. We carry out a total of $1200$ attacks. 

\begin{table}[t]
\centering
\small
\caption{Types and descriptions of attacks.}\label{table:attack}
\scalebox{0.92}{
\begin{tabular}{p{1cm}||p{5.5cm}|p{7.3cm}}
\hline Types & Attacks & Description \\
\hline 
\multirow{5}{*}{\textbf{Type 1}} & (1) modify/remove identifiers & Modifying or removing variable names, function names, and class names, etc.\\
& (2) modify/remove inputs and outputs & Modifying or removing the content of program inputs and outputs.\\
& (3) modify/remove comments & Modifying or removing single-line comments or multi-line comments.\\
& (4) modify/remove User-defined data & Modifying or removing user-customizable data, such as numbers, strings, etc. \\
& (5) modify/remove assignment statements & Modifying or removing assignment operations, such as conditional statements and comparisons.\\
\hline
\multirow{3}{*}{\textbf{Type 2}} & (6) add comments & Adding comments at any position. \\
& (7) add assignment operation  & Adding assignment statements \\
& (8) add redundant statements & Adding useless statements, example of defining variables \\
\hline
\end{tabular}
}
\end{table}

We conduct the aforementioned attacks on four watermarking strategies, resulting in a total of 4,800 attacks. The results are shown in Table~\ref{table:rq3watermarkcompare}. As observed, \textsc{MCGMark} achieved relatively favorable results across all eight types of attacks, whereas other watermarking strategies often exhibited weaker robustness under specific types of attacks. For instance, WLLM demonstrated poor robustness when faced with modifications involving longer text, such as removing or adding comments. This is because WLLM relies on statistical rules to determine whether a piece of code contains a watermark. Large-scale removal of comments, while not affecting code functionality, significantly impacts WLLM’s watermark detection performance.

Similarly, PostMark shows weaker robustness against attacks that modify or remove code elements. MPAC performs more robustly across the eight attack types, likely due to mechanisms like List Decoding, but struggles with Type 1 attacks. Overall, \textsc{MCGMark} demonstrates consistently high robustness.

 \textbf{Robustness Analysis.} \textsc{MCGMark} failed to defend against attacks, to further investigate its robustness. In $1200$ attacks, we achieve a complete defense against Attacks 3, 6, 7, and 8. However, Attacks 4 and 1 failed $13$ times and $9$ times, respectively, resulting in defense success rates of $91.3\%$ and $94\%$. Upon carefully examining the failed instances, we identify a flaw in our token matching strategy during the implementation of \textit{Pattern 1-7}. We rely on a generic regular expression matching approach, which proves inadequate for matching the tokens generated by LLMs due to their deviation from human language rules. The same phenomenon also affects the defense against Attack 2, resulting in an accuracy rate of $93.3\%$. Fortunately, this issue can be resolved by designing a more powerful matching scheme or SSP adapting the token vocabulary for the watermark.

The effectiveness of defense is comparatively reduced against Attack 5. Out of $150$ attacks of this type, there are $31$ defense failures, resulting in a success rate of $79.3\%$. We carefully examine these instances and find that $6$ of the failures were also due to matching issues. The remaining failures occur because of the maximum output limit of $400$ tokens. In such cases, the watermarking process could not be fully completed before reaching the maximum output limit. Fortunately, our watermarking design involves multiple rounds of watermark embedding, aiming to add as many watermarks as possible. As long as the watermark is added more than once, such cases have minimal impact on our detection performance. Additionally, relaxing the constraint on the maximum number of output tokens can also effectively address this issue.

\begin{table}[h]
\centering
\small
\caption{Robustness on various watermark.}\label{table:rq3watermarkcompare}
\scalebox{0.9}{
\begin{tabular}{l||c|c|c|c}
\hline \textbf{Attack} &\textbf{WLLM}  &\textbf{PostMark} & \textbf{MPAC} & \textbf{MCGMark} \\
\hline 
Type 1.(1)  &150/150  &148/150 &143/150 &142/150 \\
Type 1.(2)  &150/150  &143/150 &122/150 &139/150 \\
Type 1.(3)  &146/150  &105/150 &114/150 &150/150 \\
Type 1.(4)  &141/150  &133/150 &126/150 &137/150 \\
Type 1.(5)  &149/150  &137/150 &124/150 &129/150 \\
\hline
Type 2.(6)  &111/150  &74/150  &142/150 &150/150 \\
Type 2.(7)  &116/150  &118/150 &150/150 &150/150 \\
Type 2.(8)  &147/150  &113/150 &141/150 &150/150 \\
\hline
Total  &1110/1200  &971/1200 &1062/1200 &1147/1200 \\
\hline
\end{tabular}
}
\end{table}

Additionally, it is worth emphasizing that malicious developers may attempt to modify the code generated by LLMs. However, as long as the LLM use our watermark, the watermark information cannot be removed from the code.

\begin{tcolorbox}[title = {Answer to RQ 3:}] 
\small

\textsc{MCGMark} maintains a defense success rate of over $90\%$ against most attacks. In certain cases, \textsc{MCGMark} also experiences a failure in defense.
\end{tcolorbox}

\subsection{RQ4: Impact of Settings}
To address RQ4, we employ the controlled variable method to investigate the impact of various display parameters on the watermark embedding success rate.

\begin{figure}[h]
\begin{center}
\subfigure[The impact of value of $\alpha$.]
{\includegraphics[scale=.25]{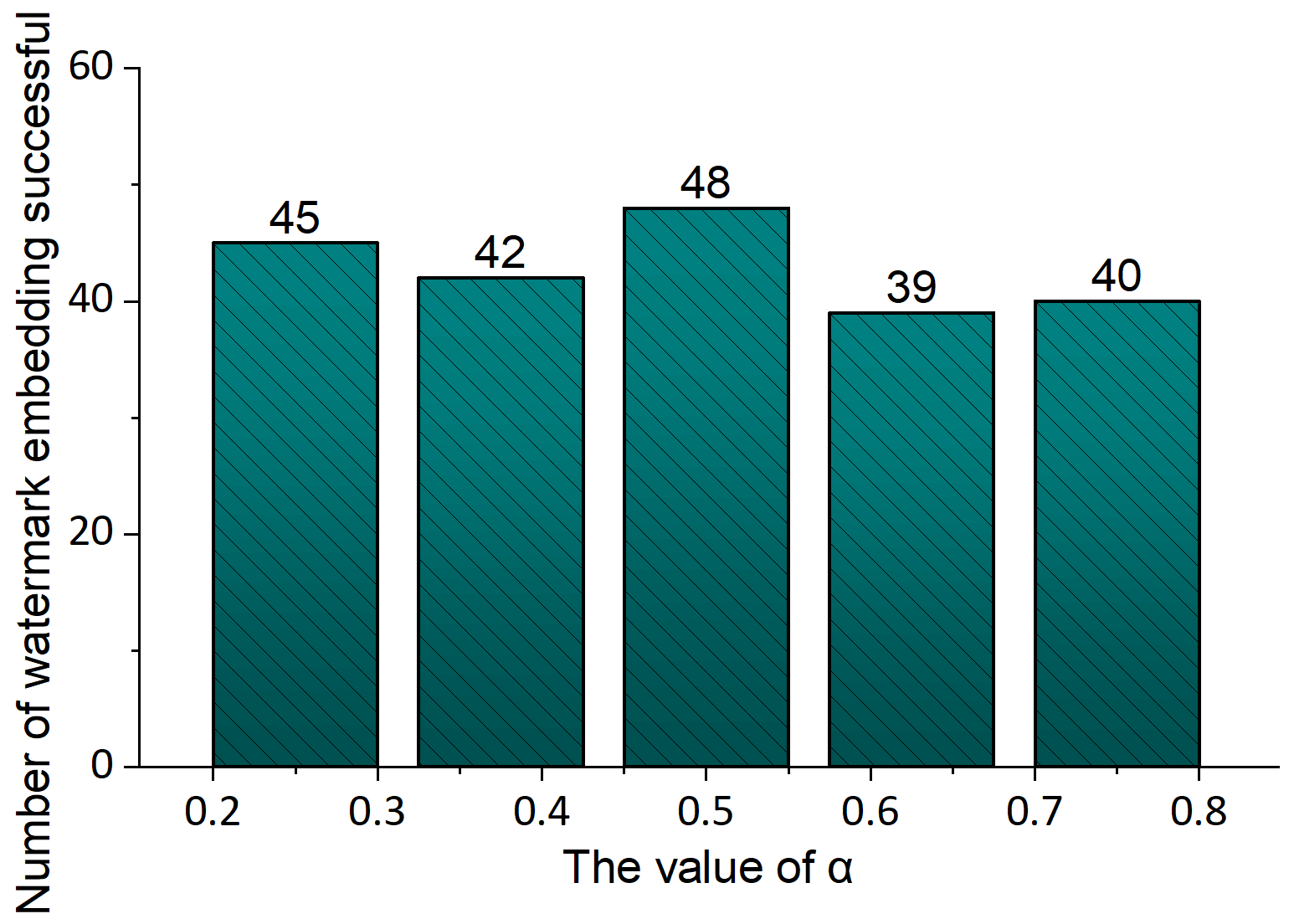}}
\subfigure[The impact of the maximum token value.]
{\includegraphics[scale=.25]{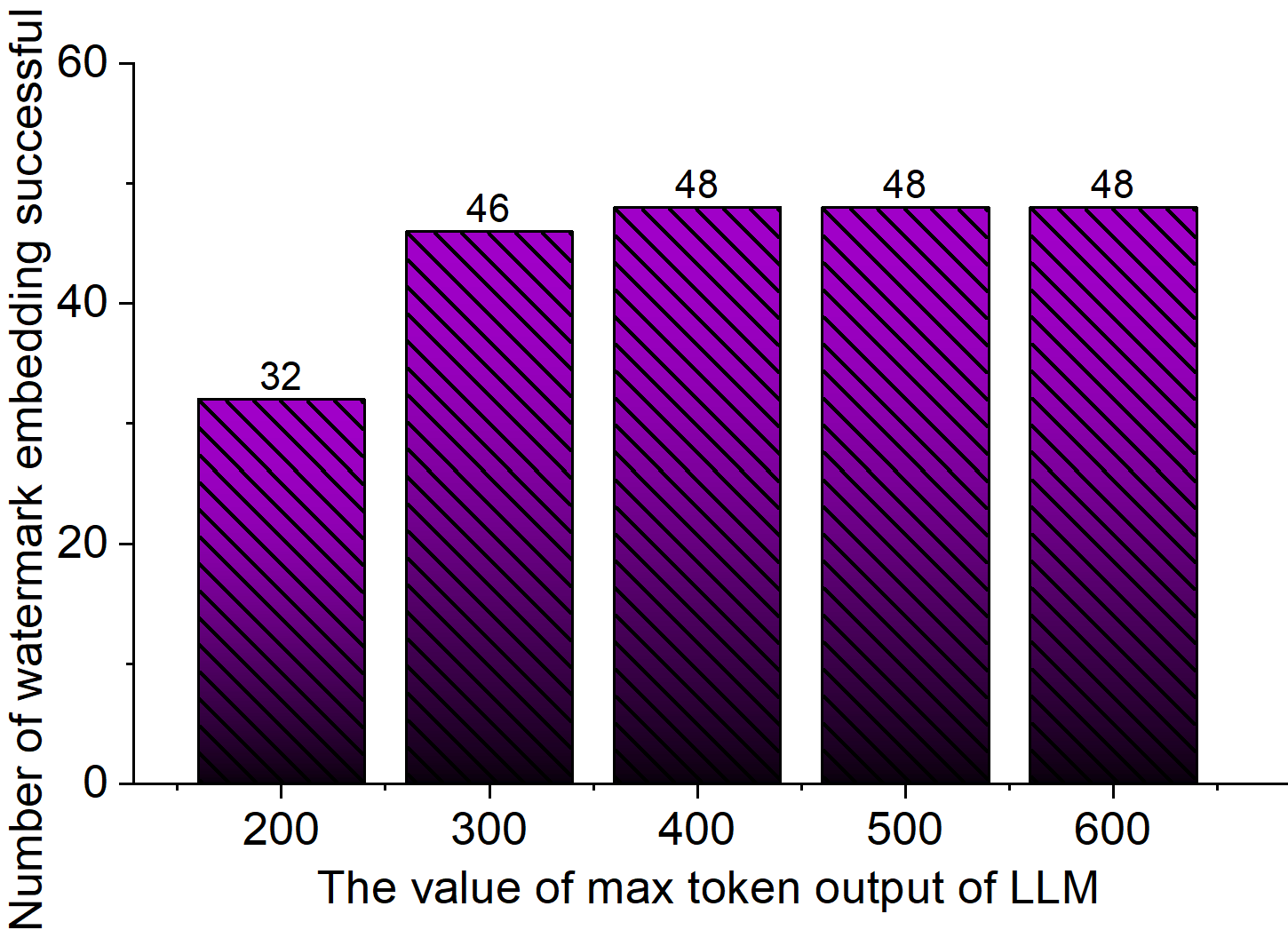}}
\caption{The impact of two hyperparameters on the success rate of watermark embedding.}
\label{fig:rq1}
\end{center}
\end{figure}

\textbf{Hyperparameters. }We examine the impact of three hyperparameters on watermark embedding success. We randomly select 50 prompts from the 406 tasks and vary one hyperparameter at a time, keeping others fixed. First, we analyze $\lceil \frac{|D|}{|V|} \rceil$, which controls the proportion of vocabulary $\mathbb{A}$ or $\mathbb{B}$ during partitioning. Following~\cite{Kirchenbauer2023watermark}, we fix the LLM’s output length at 400 and set $\lceil \frac{|D|}{|V|} \rceil$ to $[0.25, 0.375, 0.5, 0.625, 0.75]$. As shown in Fig.~\ref{fig:rq1}(a), indicate that the success rate does not increase monotonically with the vocabulary ratio. In other words, a larger vocabulary does not necessarily facilitate watermark embedding. Under the current configuration, a ratio of $0.5$ yields the best performance.

\textbf{Maximum number of output tokens. }Furthermore, we explore the impact of the maximum number of output tokens of the LLM on \textsc{MCGMark}. We keep the base setting of $\lceil \frac{|D|}{|V|}\rceil$. Simultaneously, we test the maximum number of output tokens by setting it to $[200, 300, 400, 500, 600]$. Our results are shown in  Fig.~\ref{fig:rq1}.(b). From the results, we can observe that as the maximum number of output tokens of the LLM increases, the success rate of watermark embedding also tends to increase. However, there is an evident diminishing marginal effect.

\textbf{Hash value. }Finally, we explore the impact of hash value $H$ on the watermark Embedding Success Rate. In \textsc{MCGMark}, $H$ is continuously varied to ensure the randomness of vocabulary partitioning. For comparison, we adopt a fixed-hash strategy using values $7,775$ and $666$, keeping all other settings unchanged.  Only nine watermarks were successfully embedded under the hash value of $7,775$, yielding an embedding success rate of $18\%$. Conversely, under $H$ on $666$, $19$ watermarks were successfully embedded, resulting in a success rate of $38\%$. We further test scenarios with $H$ values of $15,485,863$ and two, resulting in embedding success rates of $36\%$ and $44\%$, respectively. This phenomenon leads to two key conclusions: (1) The choice of hash values affects the success rate of watermark embedding. Since hash values are typically private to the SSP, selecting an appropriate initial hash value can further improve watermark embedding performance.
(2) \textsc{MCGMark}’s pseudo-random hash strategy proves highly effective, significantly improving the embedding success rate compared to fixed hash values. By adopting \textsc{MCGMark}’s approach, SSP can eliminate the need to spend additional time searching for optimal hash values.

\begin{table}[h]
\small
\centering
\caption{The impact of fixed hash values on watermark embedding success rate.}\
\scalebox{0.9}{
\begin{tabular}{l||c|c|c|c}
\hline \textbf{Fixed Hash Key} & 7,775 & 666 & 15,485,865 & 2 \\
\hline \textbf{Embedding Success Rate} & $18\%$ & $38\%$ & $36\%$ & $44\%$ \\

\hline
\end{tabular}
}
\end{table}

\begin{tcolorbox}[title = {Answer to RQ 4:}] 
\small
The maximum output token count of the LLM, the proportion of vocabulary partitioning $\lceil \frac{|D|}{|V|}\rceil$, and the hash value
$H$ all influence the success rate of watermark embedding. This indirectly highlights the importance of mechanisms such as reproducible floating hashes that we have implemented.
\end{tcolorbox}

\subsection{RQ5: Extra time overhead}
\label{sec:overhead}

To address RQ5, we analyze the additional time overhead introduced by \textsc{MCGMark}. Specifically, we examine the additional time overhead of \textsc{MCGMark} when applied to different LLMs and compared it with other watermarking strategies when using the same LLM.

 \textbf{Overhead with various LLM. }We evaluate the time overhead introduced by \textsc{MCGMark} on three state-of-the-art LLMs. To clearly illustrate the impact of \textsc{MCGMark}, we also measure the time overhead of the LLMs without watermarking. The evaluation was conducted on all 406 prompts from \textsc{MCGTest}, and the average results were calculated to provide a more intuitive comparison. The results are shown in Table~\ref{table:rq5watermarkcompare1}. From the results, we can observe that \textsc{MCGMark} does introduce additional time overhead, and this overhead shows some correlation with the characteristics of the respective LLMs.

\begin{table}[h]
\centering
\small
\caption{Time overhead of \textsc{MCGMark} across different LLMs. (\textit{wo}) indicates the setting without watermarking, and (\textit{w}) indicates the setting with watermarking.}
\label{table:rq5watermarkcompare1}
\scalebox{0.9}{
\begin{tabular}{l||c|c|c}
\hline \textbf{LLM} &\textbf{Overhead (\textit{wo})}  &\textbf{Overhead (\textit{w})} & \textbf{Multiple}  \\
\hline 
Deepseek-Coder  & 14.04 &103.48 & 7.37 \\
StarCoder-2 & 11.69 &120.55 & 10.31 \\
CodeLlama & 9.65 & 91.31 & 9.87 \\

\hline

\end{tabular}
}
\end{table}

 \textbf{Overhead of various watermark. }We further evaluate the time overhead by different watermarking strategies, as shown in Table~\ref{table:rq5watermarkcompare2}. To better illustrate the result, we measure both the embedding overhead and detection overhead for each watermarking strategy. From the results, we observe that all watermarking strategies introduce additional time overhead. Among them, WLLM, the online zero-bit watermarking method, is the most lightweight. In contrast, PostMark incurs higher time overhead as it requires invoking the LLM during both code generation and watermark embedding. The overall time overhead of MPAC and \textsc{MCGMark} is similar, with both methods introducing more latency during watermark embedding. Meanwhile, MPAC requires additional operations in its implementation pipelines, leading to higher embedding delay. Overall, \textsc{MCGMark} remains competitive compared to other baselines.

\begin{table}[h]
\centering
\small
\caption{Time overhead of various watermarking strategies. (\textit{E}) indicates the time required for watermark embedding, (\textit{D}) indicates the time for watermark detection, and (\textit{T}) represents the total overhead.}
\label{table:rq5watermarkcompare2}
\scalebox{0.9}{
\begin{tabular}{l||c|c|c}
\hline \textbf{Watermark} &\textbf{Overhead (\textit{E})}  &\textbf{Overhead (\textit{D})} & \textbf{Overhead (\textit{T})}  \\
\hline 
No watermark  & / & /  & 14.04 \\
WLLM & 41.62 & 1.29 & 42.91 \\
PostMark & 202.48 & 0.21 & 202.69 \\
MPAC & 112.31 & 2.14 & 114.45 \\
MCGMark & 99.12 & 4.36 & 103.48 \\
\hline

\end{tabular}
}
\end{table}

\begin{tcolorbox}[title = {Answer to RQ 5:}] 
\small
\textsc{MCGMark} introduces a measurable level of additional time overhead. However, it still achieves competitive results compared to the baselines.
\end{tcolorbox}

\section{Limitations}
\noindent\textbf{Defense Against Specific Attacks. }In Section~\ref{sec:Exper-robust}, the experiments demonstrate that the watermark exhibits overall robustness and can withstand most tampering attacks, but it may fail in certain specific cases. Analysis shows incomplete token adaptation due to reliance on regex-based recognition of extensive LLM vocabularies (over 3.2K tokens) as a primary issue. Improving token adaptation through refined recognition techniques and SSP-based support constitutes a direction for future research. Additionally, inserting non-watermarked human-written code can contaminate detection, though this typically changes the code’s functionality, making it effectively non-LLM-generated. Enhanced watermark truncation and extraction techniques are considered promising directions for future studies to address such attacks.

\noindent\textbf{Inevitable Impact on Code Quality. }Another limitation is the impact of the watermark on the quality of LLM output. To address this, we design \texttt{Algorithm~\ref{Alg:Alg1}} based on probabilistic outliers in the LLM's vocabulary to ensure optimal token selection. However, the watermark inevitably impacts code quality. This occurs for two reasons: first, outliers themselves might have quality variations; second, the error correction bit of the watermark can influence the code output. Fortunately, the criteria for filtering outliers can be adjusted flexibly. A looser outlier filtering standard ensures more random model output, while a stricter standard guarantees higher output quality. Further compressing the length of the watermark, especially the error correction bit, can lead to even greater reductions in impact.

\noindent\textbf{Unavoidable Overhead Introduction. }A further limitation of \textsc{MCGMark} is the additional time overhead, a common issue for all watermarking schemes~\cite{zhang2023watermarks,zhang2024watermarking}. Nevertheless, as shown in Section~\ref{sec:overhead}, its overall overhead remains comparable to the baselines. \textsc{MCGMark} also avoids loading extra models or databases during embedding and requires only the malicious code for detection, without relying on the LLM, external resources, or intermediate states. This design improves space efficiency and practicality for real-world SSP traceability. Future work will focus on optimizing hash computation and vocabulary partitioning to further reduce time and space complexity.

\section{Threats to validity}

\subsection{Internal Validity}

\noindent\textbf{Subjective Bias in Manual Analysis. }In section~\ref{sec:promptdataset}, the construction of the \textsc{MCGTest} relied on manual analysis by participants, which introduces a certain degree of subjectivity to the results. To mitigate this threat, we use a close card sorting method for each aspect requiring manual analysis, involving at least two participants to ensure consistency in the results. We make significant efforts to mitigate this potential threat.

\noindent\textbf{Limitations of Data Collection Methods. }Another internal validity concern in \textsc{MCGTest} is its reliance on keyword matching for data collection, which may be incomplete. However, the goal is to gather sufficient malicious code scenarios to design, test, and advance \textsc{MCGMark}. Rather than capturing every instance, we focus on representative cases that meet our research needs.

\noindent\textbf{Scale Limitations of Evaluation Benchmarks. }Additionally, we acknowledge that \textsc{MCGMark} was evaluated on the \textsc{MCGTest}, which may have scale limitations. However, with 406 tasks, \textsc{MCGTest} is already substantial. In comparison, widely used LLM benchmarks like OpenAI’s HumanEval~\cite{chen2021evaluatinglargelanguagemodels} contain only 164 tasks. Moreover, as noted by CodeIP~\cite{guan2024codeip}, benchmarks such as HumanEval and MBPP~\cite{roziere2023codellama} focus on simple problems with short generated code, making them unsuitable for assessing longer multi-bit watermark embedding.

\subsection{External Validity }

\noindent\textbf{Scope of Model Adaptability. }We only evaluated \textsc{MCGMark} on three open-source LLMs, without extending the assessment to additional models. However, the watermark embedding, detection, and robustness algorithms in  \textsc{MCGMark} are not tailored specifically to these evaluated LLMs. They are designed to be applicable to any LLM based on the transformer architecture~\cite{kirchenbauer2023reliability}. Adapting \textsc{MCGMark} to other models involves adjusting the code element matching in \texttt{Algorithm~\ref{Alg:Alg2}} to fit the model’s vocabulary. This adjustment is straightforward and does not present significant technical barriers.

\noindent\textbf{Generalizability Across Programming Languages. }In this paper, watermark patterns are specifically tailored to Python due to its current dominance among attackers~\cite{acarturk2021malicious}. However, extending these watermark patterns to other programming languages does not present substantial technical challenges. The process requires only minor modifications to map and match the target language’s code elements to our established watermark patterns. It is worth noting that because our watermark embedding process operates synchronously with code generation, we do not have access to the complete source code at embedding time. Thus, extending \textsc{MCGMark}’s adaptability using AST-based approaches is not feasible in the current design.

\section{Discussion}

While evaluating \textsc{MCGMark}, we identify several scenarios that pose challenges to current watermarking techniques. This section discusses these scenarios and proposes potential solutions to address them. 

\subsection{Watermark Embedding}

\noindent\textbf{Short Code Generation. }Embedding multi-bit watermarks while ensuring robustness is particularly challenging in short code generation scenarios. Although this paper tests with 400 tokens (DeepSeek-Coder supports up to 2048 tokens, while models like Codellama generate sequences of up to 100,000 tokens), \textsc{MCGMark} struggles in extremely short code scenarios. Designing multi-bit watermarks for shortcode generation without compromising robustness remains a significant challenge. One potential solution involves compressing watermark encoding using higher-dimensional vocabulary partitions.

\noindent\textbf{Poor Model Outputs.} Watermark embedding often fails when model outputs are subpar, such as generating syntactically incorrect code or mixing natural language with code. Online watermarking relies heavily on the LLM’s generation capabilities, making a more powerful model the most direct solution. Additionally, embedding strategies should incorporate rollback mechanisms based on generated tokens to preserve output quality.

\subsection{Watermark Detection}

\noindent\textbf{Tokenization Discrepancies.} Tokenizers are not always reliable. Minor discrepancies between tokenization during code generation and code detection may lead to errors in watermark detection. Notably, if tokens affected by these discrepancies do not contain watermarks, automatic correction may occur during subsequent detection processes. Conducting systematic empirical studies on tokenizer errors could provide insights into addressing this issue. Additionally, due to its reliance on tokenizers, \textsc{MCGMark} is only applicable for detecting and tracing code generated by LLMs. It is not suitable for evaluating human-written code, as such code may not be accurately tokenized, potentially leading to unexpected results. Furthermore, the evaluation of human-written code falls outside the scope of this work.

\noindent\textbf{Watermark Strength.} The strength of the watermark directly affects the success rate of watermark detection; lower strength results in detection failures, while excessively high strength impacts the output quality of the model. To mitigate this issue, this paper proposes \texttt{Algorithm~\ref{Alg:Alg1}} and designs a watermark prompt. However, this approach still affects the output quality to some extent. Balancing watermark strength and detection success rate remains a significant challenge worth exploring.

\section{Related work}

\subsection{Traditional Code Watermark} 

Code watermarking involves directly adding a special identifier to the source code or during code execution to declare code ownership~\cite{kitagawa2022watermarking}. Code watermarking techniques can be broadly categorized into static and dynamic approaches~\cite{li2023zhengju2}. Static watermarking embeds watermarks directly into the source code. For instance, Kim et al.~\cite{kim2023smartmark} uses adaptive semantic-preserving transformations to embed watermarks, and Sun et al.~\cite{sun2023codemark} does so by changing the order of functions. However, static watermarks are relatively more susceptible to detection and removal~\cite{softmark}. Consequently, dynamic code watermarking techniques have seen rapid development recently. For example, LLWM~\cite{novac2021llwm} is a watermarking technique that uses LLVM and Clang to embed watermarks by compiling the code. Xmark~\cite{ma2019xmark} embeds watermarks by obfuscating the control flow based on the Collatz conjecture. However, dynamic watermarking techniques are not applicable to the code generation process of LLMs, as LLMs do not execute the generated code.

\textbf{Difference. }Therefore, current traditional code watermark are not suitable for LLM code generation. The watermarking techniques mentioned above differ significantly from the watermark proposed in this paper, which operates during the LLM code generation process. We need to design more innovative watermarks for the LLM code generation task.

\subsection{LLM Watermark}

The security of LLMs has recently attracted significant attention, leading to the adoption of watermarking techniques for protection. Watermarking aims to prevent model theft, particularly through distillation in model security. For instance, GINSW~\cite{zhao2023protecting} embeds private signals into probability vectors during decoding to deter theft. TOSYN~\cite{LiWWG23} replaces training samples with synthesized code to defend against distillation attacks. PLMmark~\cite{li2023plmmark} embeds watermarks during LLM training to establish model ownership.

\textbf{Difference. }The watermarking techniques proposed in above works can protect model copyright in various scenarios. However, they are not suitable for verifying text generated by LLMs. Therefore, there is a fundamental difference between these techniques and the problem addressed by the watermarking method proposed in this paper.

\subsection{Watermarking for LLM Text Generation}

\textbf{Non-encodable Text Watermark.} LLM watermarking techniques can be categorized into offline and online watermarking. In online watermarking, the watermarking process is synchronized with the LLM content generation process~\cite{abs2023Rela}. On the other hand, offline watermarking requires processing the generated text after the LLM has completed content generation~\cite{PengYWWZLJXSX23,chang2024postmark}. Offline watermarking is not directly related to the LLM itself. They rely more on rules and are easier for attackers to detect~\cite{liu2023survey}. Kirchenbauer et al.~\cite{Kirchenbauer2023watermark} is a text watermarking technique that has received considerable attention. It involves partitioning the vocabulary of LLM and guiding the model to select words from a predefined vocabulary, thereby embedding watermarks into the generated text~\cite{abs2023Rela}. A follow-up work~\cite{abs2023Rela} investigates the reliability of this watermarking strategy. Subsequent works have extended this watermarking technique to privacy~\cite{liu2024unforgeable} and robustness~\cite{yoo2023robust}.

\textbf{Difference. }However, this watermarking strategy can only obtain binary results. Recently, some research has focused on designing encodable watermarks for LLMs.

\noindent\textbf{Encodable Text Watermark.} Yoo et al.~\cite{yoo2024advancing}, also building on the work of Kirchenbauer et al.~\cite{Kirchenbauer2023watermark}, achieved multi-bit watermark embedding by dividing the vocabulary into more sub-vocabularies. Wang et al.~\cite{wang2024codable} analyzed how to embed more information into watermarks through vocabulary partitioning from a mathematical perspective. However, further subdividing the vocabulary may lead to a significant decline in the quality of the model's output. Boroujeny et al.~\cite{boroujeny2024multi} proposed a multi-bit watermarking embedding scheme that does not alter the probability distribution of the vocabulary but instead controls the token selection offset of the LLM.

\textbf{Difference. }The aforementioned works are only applicable to text-generation scenarios. Code, as a special type of text, has a relatively fixed structure and pattern. These watermarks are not designed for code generation scenarios and cannot address the issues proposed in this paper.

\subsection{Watermarking for LLM Code Generation}

As the issue of malicious developers using LLMs to generate malicious code becomes more recognized, some works have designed watermarking schemes specifically for LLM code generation. Li et al.~\cite{li2024resilient} devised a set of code transformation rules to embed watermarks through post-processing. Yang et al.~\cite{yang2024srcmarker} designed an AST-based code transformation method to embed multi-bit watermarks, which is also a post-processing watermark. Post-processing watermarks are more susceptible to attackers discovering their rules and disrupting the watermark. Additionally, neither of these methods offers robust solutions tailored to the structure of the code. SWEET~\cite{lee2024zhengju3} is an online code watermark that extends the low-entropy scenarios of Kirchenbauer et al.~\cite{Kirchenbauer2023watermark} to improve code generation quality. However, SWEET can only achieve binary results and still does not address the issues proposed in this paper. CodeIP~\cite{guan2024codeip} designed a code-based multi-bit watermarking scheme, which restricts the sampling process of predicting the next token by training a type predictor.

\textbf{Difference. }The approach proposed in this paper is fundamentally different from any of the above methods. It is an online watermarking scheme, making the watermark more challenging to detect and remove. Furthermore, our watermark is encodable and capable of embedding the creator's identity information. It also enhances robustness against code structure to prevent malicious developers from easily breaking the watermark through simple modifications.

\section{Conclusion}
In this work, we propose \textsc{MCGMark}, a robust and encodable watermarking technique for LLMs to counteract the growing trend of malicious code generation. \textsc{MCGMark} embeds user identity information implicitly into the generated code by controlling the LLM’s token selection process. To improve watermark quality, \textsc{MCGMark} leverages probabilistic outliers in the vocabulary to optimize the quality of candidate tokens. Additionally, to enhance robustness, it uses code structure and syntax rules to skip easily modifiable elements such as comments, reducing the risk of watermark removal. Furthermore, we construct \textsc{MCGTest}, the first prompt dataset targeting LLM-generated malicious code. It consists of 406 tasks covering both real-world instances and potential risk scenarios. We conduct a comprehensive evaluation of \textsc{MCGMark} on this dataset, and the results demonstrate that it achieves strong performance in terms of embedding success rate, generation quality, robustness, and time efficiency. We have open-sourced both \textsc{MCGMark} and \textsc{MCGTest} to provide a reproducible foundation for the research community. In future work, we plan to explore watermark compression techniques to further reduce interference with code generation, and develop more general token-matching mechanisms to support complex token structures across different LLM architectures, thereby improving the generality and defense capability of the watermark.


\balance
    \bibliographystyle{ACM-Reference-Format}
    \bibliography{ref}

\end{document}